\documentclass[12pt]{article}
\pdfoutput=1

\usepackage[usenames,dvipsnames,svgnames,table]{xcolor} 
\usepackage[obeyspaces,hyphens,spaces]{url}
\usepackage{jcapmod}
\usepackage{verbatim}
\usepackage{graphics}
\usepackage{epsfig}
\usepackage{booktabs}
\usepackage{booktabs}
\usepackage[english]{babel}
\usepackage{amsmath, amssymb, amsbsy, amstext, amsthm, simplewick}
\usepackage{hyperref}
\usepackage{graphicx}
\usepackage{amsfonts}
\usepackage{rotating}
\usepackage{upgreek}
\usepackage{framed}
\usepackage{tensor}
\usepackage{placeins}

\let\Oldsection\section
\renewcommand{\section}{\FloatBarrier\Oldsection}

\let\Oldsubsection\subsection
\renewcommand{\subsection}{\FloatBarrier\Oldsubsection}

\let\Oldsubsubsection\subsubsection
\renewcommand{\subsubsection}{\FloatBarrier\Oldsubsubsection}

\usepackage{pifont}
\usepackage{latexsym, mathrsfs}
\usepackage{array}
\usepackage{hyperref}
\usepackage{xspace}
\usepackage{longtable}
\usepackage{multirow}
\usepackage{cases}
\usepackage{empheq}

\usepackage{bm}
\usepackage{bbm}
\usepackage{float}
\usepackage{afterpage}
\usepackage{setspace}
\usepackage{caption}
\usepackage[load-configurations=astronomy, range-units=brackets, range-phrase=-, per-mode=reciprocal, mode=math]{siunitx}
\usepackage{threeparttable}
\usepackage{subfigure}
\usepackage{tcolorbox}
\usepackage{geometry}

\allowdisplaybreaks

\usepackage{braket}

\usepackage{ragged2e}

\usepackage{hhline}
\usepackage{array}
\newcolumntype{P}[1]{>{\centering\arraybackslash}p{#1}}
\usepackage{booktabs}
\usepackage{tikz}
\usetikzlibrary{calc,shadings,patterns,tikzmark,fadings}

\usepackage{cleveref} 
\Crefname{equation}{Eq.}{Eqs.}
\Crefname{section}{Sec.}{Secs.}

\Crefname{figure}{Fig.}{Figs.}
\Crefname{table}{Table}{Tables}

\newcolumntype{C}[1]{>{\centering\let\newline\\\arraybackslash\hspace{0pt}}m{#1}}

\definecolor{Blue}{rgb}{0.25, 0.41, 0.88}
\definecolor{Red}{rgb}{0.92,0.,0.}
\definecolor{darkorange}{rgb}{1.0,0.549,0.}
\definecolor{cobalt}{RGB}{44, 98, 120}
\definecolor{Mathematica1}{rgb}{0.368417, 0.506779, 0.709798}
\definecolor{Mathematica2}{rgb}{0.880722, 0.611041, 0.142051}
\definecolor{Mathematica3}{rgb}{0.560181, 0.691569, 0.194885}
\definecolor{Mathematica4}{rgb}{0.922526, 0.385626, 0.209179}
\definecolor{Mathematica5}{rgb}{0.528488, 0.470624, 0.701351}
\definecolor{Mathematica6}{rgb}{0.772079, 0.431554, 0.102387}
\definecolor{Mathematica7}{rgb}{0.363898, 0.618501, 0.782349}
\definecolor{Mathematica8}{rgb}{1, 0.75, 0}
\definecolor{Mathematica9}{rgb}{0.647624, 0.37816, 0.614037}
\definecolor{plotBlue}{RGB}{94, 130, 181}
\definecolor{plotRed}{RGB}{233, 85, 54}
\definecolor{plotGreen}{RGB}{142, 176, 50}
\definecolor{plotPurple}{RGB}{135, 120, 178}

\newcolumntype{C}[1]{>{\centering\let\newline\\\arraybackslash\hspace{0pt}}m{#1}}

\makeatletter
\newlength{\apb@width}
\newcommand{\autoparbox}[2][c]{\settowidth{\apb@width}{#2}\parbox[#1]{\apb@width}{#2}}

\makeatother

\makeatletter
\newsavebox\myboxA
\newsavebox\myboxB
\newlength\mylenA

\newcommand*\xoverline[2][0.75]{
	\sbox{\myboxA}{$\m@th#2$}
	\setbox\myboxB\null
	\ht\myboxB=\ht\myboxA
	\dp\myboxB=\dp\myboxA
	\wd\myboxB=#1\wd\myboxA
	\sbox\myboxB{$\m@th\overline{\copy\myboxB}$}
	\setlength\mylenA{\the\wd\myboxA}
	\addtolength\mylenA{-\the\wd\myboxB}
	ifdim\wd\myboxB<\wd\myboxA
	\rlap{\hskip 0.5\mylenA\usebox\myboxB}{\usebox\myboxA}
	\else
	\hskip -0.5\mylenA\rlap{\usebox\myboxA}{\hskip 0.5\mylenA\usebox\myboxB}
	\fi}
\makeatother

\numberwithin{equation}{section}
\numberwithin{figure}{section}
\numberwithin{table}{section}

\def\beq{\begin{equation}}
\def\eeq{\end{equation}}

\def\bea{\begin{eqnarray}}
\def\eea{\end{eqnarray}}

\def\beq{\begin{equation}}
\def\eeq{\end{equation}}
\def\bea{\begin{eqnarray}}
\def\eea{\end{eqnarray}}

\numberwithin{equation}{section}
\def\beq{\begin{equation}}
\def\eeq{\end{equation}}

\def\bea{\begin{eqnarray}}
\def\eea{\end{eqnarray}}

\DeclareRobustCommand{\SkipTocEntry}[4]{}

\usepackage{colortbl}
\definecolor{blue2}{cmyk}{1, 0.1, 0.1, 0.1}

\definecolor{pyBlue}{RGB}{31, 119, 180}
\definecolor{pyRed}{RGB}{214, 39, 40}
\definecolor{pyGreen}{RGB}{44, 160, 44}
\definecolor{pyBlue2}{RGB}{0, 111, 237}
\definecolor{pyRed2}{RGB}{224, 52, 36}

\newcolumntype{P}[1]{>{\centering\arraybackslash}p{#1}}
\newcolumntype{M}[1]{>{\centering\arraybackslash}m{#1}}

\begin{document}
	
	\tcbset{colframe=black,arc=0mm,box align=center,halign=left,valign=center,top=-10pt}
	
	\renewcommand{\thefootnote}{\fnsymbol{footnote}}
	
	\pagenumbering{roman}
	\begin{titlepage}
		\baselineskip=5.5pt \thispagestyle{empty}
		
		\bigskip\
		
		\vspace{0.1cm}
		\begin{center}
			
			{\Huge \textcolor{Sepia}{\bf \sffamily ${\cal C}$}}{\large \textcolor{Sepia}{\bf\sffamily IRCUIT}} ~{\Huge \textcolor{Sepia}{\bf \sffamily ${\cal C}$}}{\large \textcolor{Sepia}{\bf\sffamily OMPLEXITY}}
			~\vspace{0.45cm} \\~{\Huge \textcolor{Sepia}{\bf \sffamily ${\cal F}$}}{\large \textcolor{Sepia}{\bf\sffamily ROM}} \vspace{0.45cm} \\{\Huge \textcolor{Sepia}{\bf \sffamily ${\cal S}$}}{\large \textcolor{Sepia}{\bf\sffamily UPERSYMMETRIC}} {\large \textcolor{Sepia}{\bf\sffamily }}~ {\Huge \textcolor{Sepia}{\bf\sffamily ${\cal Q}$}}{\large \textcolor{Sepia}{\bf\sffamily UANTUM }}~ {\Huge \textcolor{Sepia}{\bf\sffamily ${\cal F}$}}{\large \textcolor{Sepia}{\bf\sffamily IELD}}~
		{\Huge \textcolor{Sepia}{\bf\sffamily ${\cal T}$}}{\large \textcolor{Sepia}{\bf\sffamily HEORY }} \vspace{0.45cm}\\
		~ {\Huge \textcolor{Sepia}{\bf\sffamily ${\cal W}$}}{\large \textcolor{Sepia}{\bf\sffamily ITH }}  \vspace{0.45cm}\\
		~ {\Huge \textcolor{Sepia}{\bf\sffamily ${\cal M}$}}{\large \textcolor{Sepia}{\bf\sffamily ORSE }}~ {\Huge \textcolor{Sepia}{\bf\sffamily ${\cal F}$}}{\large \textcolor{Sepia}{\bf\sffamily UNCTION }}

		\end{center}
		
		\vspace{0.05cm}

			\begin{center}

				{\fontsize{15}{18}\selectfont Sayantan Choudhury${}^{\textcolor{Sepia}{1},{2}}$\footnote{{\sffamily \textit{ Corresponding author, E-mail}} : {\ttfamily sayantan.choudhury@niser.ac.in, sayanphysicsisi@gmail.com}}}${{}^{,}}$
				\footnote{{\sffamily \textit{ NOTE: This project is the part of the non-profit virtual international research consortium ``Quantum Aspects of Space-Time \& Matter" (QASTM)} }. }, 
				{\fontsize{15}{18}\selectfont Sachin Panneer Selvam${}^{\textcolor{Sepia}{3}}$},
{\fontsize{15}{18}\selectfont K. Shirish${}^{\textcolor{Sepia}{4}}$},
				
			\end{center}

		\begin{center}
			\vskip4pt
		{
				
				\textit{${}^{1}$National Institute of Science Education and Research, Bhubaneswar, Odisha - 752050, India}\\
		\textit{${}^{2}$Homi Bhabha National Institute, Training School Complex, Anushakti Nagar, Mumbai - 400085, India}\\	
		\textit{${}^{3}$ Department of Physics, Birla Institute of Technology and Science, Pilani, Hyderabad Campus, Hyderabad - 500078, India}\\			
				\textit{${}^{4}$Visvesvaraya National Institute of Technology, Nagpur, Maharashtra, 440010, India}\\
			}
		\end{center}
		
		\vspace{0.2cm}
		\hrule 
		\begin{center}
			\textbf{Abstract}
		\end{center}
		Computation of circuit complexity has gained much attention in the Theoretical Physics community in recent times to gain insights into the chaotic features and random fluctuations of fields in the quantum regime. Recent studies of circuit complexity take inspiration from Nielsen's geometric approach, which is based on the idea of optimal quantum control in which a cost function is introduced for the various possible path to determine the optimum circuit. In this paper, we study the relationship between the circuit complexity and Morse theory within the framework of algebraic topology, which will then help us study circuit complexity in supersymmetric quantum field theory describing both simple and inverted harmonic oscillators up to higher orders of quantum corrections. We will restrict ourselves to $\mathcal{N} = 1$ supersymmetry with one fermionic generator $Q_{\alpha}$. The expression of circuit complexity in quantum regime would then be given by the Hessian of the Morse function in supersymmetric quantum field theory. We also provide technical proof of the well known universal connecting relation between quantum chaos and circuit complexity of the supersymmetric quantum field theories, using the general description of Morse theory.

		\noindent

		\vskip10pt
		\hrule
		\vskip10pt
		
		\noindent
		\text{Keywords: Circuit Complexity,  Supersymmetric QFT,  Morse Theory, AdS/CFT.} 
		
	\end{titlepage}

\newpage

\tableofcontents

\newpage

	\clearpage
	\pagenumbering{arabic}
	\setcounter{page}{1}
	
	\renewcommand{\thefootnote}{\arabic{footnote}}

\textcolor{Sepia}{\section{\sffamily Introduction}\label{sec:introduction}}
 AdS/CFT correspondence has helped in providing great insights on the geometry of the bulk from information in the boundary CFT \cite{Lashkari:2013koa, VanRaamsdonk:2010pw, Maldacena:2013xja, Maldacena:1997re, Li:2017hdt, Choudhury:2017tax}. \textcolor{blue}{In essence it means, that given a conformal field theory is holographic and it has a dual theory in classical gravity, then the states of these CFTs are associated with a certain state of the dual theory - an empty vacuum corresponds to a pure AdS, a thermal state on $S^d$ to a Scwarzschild Blach hole etc. The Ryu-Takayanagi formula interprets the entropy of any CFT substate, for a CFT with a gravitational dual. Generalizations of the formula \cite{nakata2021new,barbon2021generalized,yang2022kind,susskind2021entanglement} have been used in various descriptions to study bulk aspects of gravitational systems especially black-holes using holographic entanglement and information of the boundary CFT.} \cite{emparan2022holographic,bhattacharya2021islands,jiang2020holographic,auzzi2020subregion,an2018time}
 However, probing the physics behind the horizon is still a major challenge. It has been inspected that even if the entangled entropy of an eternal AdS black hole saturates after reaching the equilibrium, the size of the Einstein-Rosen(ER) bridge continues to grow with time, posing a problem to the dual description of boundary CFT. Considering this, Susskind has introduced new observables in bulk geometry \cite{Susskind:2014moa, Susskind:2014jwa, Shenker:2013pqa, Swingle:2014uza, Choudhury:2017bou, Choudhury:2017qyl, Choudhury:2016cso}. \textcolor{blue}{These new observables are measures of quantum information that assist in further probing the wormhole} \cite{Belin:2021bga}. One is the volume of a maximal co-dimension-one bulk surface extending to the boundary of AdS space-time, and the second is the action defined on the Wheeler-De-Witt patch. According to the conjecture, these new observables are a dual description of the complexity of the boundary field theory.
 
 One of these crucial observables is \textit{volume},  which according to ``\textit{Complexity = Volume}'' conjecture,  states that volume $\mathcal{V}(\mathcal{B})$ of a maximal co-dimension-one bulk surface $\mathcal{B}$ that extends to the AdS boundary and asymptotic to the time slice $\sum$ is proportional to the complexity of the boundary state, $C_{\mathcal{V}}\left(\sum\right)$, which is given by:
\begin{equation}
\large{ C_{\mathcal{V}}\left(\sum\right)} =  max\left[\frac{\mathcal{V(\mathcal{B})}}{G_{N}l}\right].
\end{equation} 

 The second observable is the \textit{gravitational action} estimated in the Wheeler-De-Witt (WDW) patch in the bulk region $\mathcal{I}_{WDW}$, which according to the another important conjecture, namely the      ``\textit{Complexity = Action}'' conjecture; is proportional to complexity of the boundary field theory \cite{Roberts:2014isa}, given by:
 \begin{equation}
 \large{ C_{A}=\frac{\mathcal{I}_{WDW}}{\pi\hbar}}.
 \end{equation}
 
Complexity in quantum field theory and various other quantum system has been the attraction of the theoretical physics community in recent times. \textcolor{blue}{The growth of quantum complexity in many systems and similar behaviour inside black holes has given insights about what happens inside the horizon of a gravitational dual and helps understanding it in terms of entanglement of the boundary fields} \cite{Brown:2015lvg,Brown:2015bva}. \textcolor{blue}{Recent works have shown that the ambiguties in formulation of quantum complexity (like the choice of operations/gates) is highly sensitive and a class of these observables exist and can be used to study information theoretic properties of the bulk and even relate it to the classical entropy as their growth rates are very similar} \cite{Belin:2021bga, Brown:2017jil}. \textcolor{blue}{Circuit complexity has been used in the context of many field theories in recent times to study decaying and scattering phenomenon outside equlibrium} \cite{chagnet2022complexity,bhattacharyya2021circuit,guo2018circuit,chapman2018toward} In other words, it helps in understanding the out-of-time-order-correlation (OTOC) function of observables in a quantum system that depicts the chaotic and random quantum fields \cite{Akhtar:2019qdn,Banerjee:2021lqu,Krishnan:2021faa}. OTOC within the framework of supersymmetric quantum mechanics was studied in the simple harmonic oscillator, one dimensional infinite potential well and various other models in \cite{Bhagat:2020pcd,Choudhury_2021c}. The main idea was to put the free scalar field theory on a lattice, which reduces it to the family of coupled oscillators, then identifying the circuit as a path ordered exponential of Hamiltonian, which forms the representation of $GL(2, R)$, and then construct the Euclidean metric. Minimizing the length of this metric would finally give the expression for circuit complexity. These results were obtained in the inverted harmonic oscillator framework and interacting field theories to look for chaotic behaviour in quantum field theories(QFT). In this paper, our main objective is to calculate circuit complexity for supersymmetric quantum field theories for simple and inverted harmonic oscillators for various higher-order quantum corrections. We will restrict ourselves to $\mathcal{N} = 1$ supersymmetry, such that we will have only one fermionic annihilation and creation operator and thereby only one supercharge. We will take an unusual approach and first try to connect complexity in quantum field theories, namely, supersymmetric field theories, with the Morse function. Morse theory acts as an essential tool to study the topology of manifolds by studying the differentiable functions through which we could identify the critical points on that manifold: the minima, maxima, and saddle points. To study the connection between circuit complexity in supersymmetric quantum field theories with the Morse function, we will first identify the so-called ``cost function", an important parameter in computing the circuit complexity. Cost function counts the number of gates operating at a particular time $t$ to construct the optimal circuit with the Morse function on a given manifold. By computing the \textit{Hessian} of the Morse function, we obtain the expression for circuit complexity in the present context of the discussion. Our motivation for doing so lies in the fact that the eigenvalues of supersymmetric Hamiltonian are closely concentrated near the Morse function's critical points. The number of zero eigenvalues of the ground state is exactly equal to the Betti number of the manifold. In this paper, we will show how the increase in the number of critical points of the manifold captures the amount of chaos present in supersymmetric quantum field theories for the inverted harmonic oscillator (IHO) and how the mathematical structure of supersymmetry (SUSY) in the regime inverted harmonic oscillators whose potentials forms the generators of $SL(2, R)$ group underlies chaos. It makes the number of critical points increase by a factor of exponential with respect to the superpotential appearing in supersymmetric quantum field theory. For the scalar field theories in the regime where it can be described as a simple harmonic oscillator (SHO), we will identify the non-dynamical auxiliary field; namely the $F$ - term in the lagrangian of SUSY theory involving scalar fields; with the gradient of the Morse function, which passes through every critical point on the manifold. In turn, we will show that the complexity of supersymmetric quantum field theories for simple harmonic oscillator only depends on the absolute value of the non-dynamical auxiliary field, which also acts as an order parameter for supersymmetry breaking. This way of identifying the $F$ - term with the gradient of the Morse function provides another way to check when the supersymmetry is broken. If the gradient flow passes through the critical points, then it could be said that there exist no ground states with zero energy or SUSY is spontaneously broken. Following this, we will compute the Hessian in the space of super-coordinates and will derive the expression for complexity. Edward Witten has shown that supersymmetric quantum field theory is a Hodge Derham cohomology and has derived the Morse inequality using the formalism of supersymmetry. This paper will not work along the same lines, but our main objective would be to give quantum chaos a topological flavour in supersymmetric theories. At present, Out-of-time ordered correlation (OTOC) function has been an essential theoretical tool to capture the effect of chaos in any quantum system at late time scales. By following this fundamental notion, in this paper, we will explicitly give a technical proof of the \textit{universal relation} relating circuit complexity with the quantum chaos in terms of the previously mentioned Out-of-Time Ordered Correlation Function (OTOC) within the framework of the supersymmetric quantum field by using the general description of Morse theory.
\\
\\
\textcolor{Black}{Organization of the paper is as follows:}
\\
\begin{itemize}
	\item {In \underline{section~\ref{sec:2}}, we provide a brief review of the concept of circuit complexity in the general context of quantum information theory.}
	
	\item {In \underline{section~\ref{sec:3}}, we will explain the Lie algebra formulation and how the potential of Inverted Harmonic Oscillator (IHO) could be embedded in the structure of the manifold.}
	
	\item{In \underline{section~\ref{sec:4}}, we will give a brief review of Morse theory, namely the Morse function and the gradient flow of the Morse function for completeness.}
	
	\item {In \underline{section~\ref{sec:5}}, we will comment on the relationship between circuit complexity and Morse function over a manifold in supersymmetric quantum mechanics.}
	
	\item{In \underline{section~\ref{sec:6}}, we will explicitly compute the expression of circuit complexity for the Inverted Harmonic Oscillator (IHO) up to higher-order of quantum corrections.}
	
	\item{In \underline{section~\ref{sec:7}}, we will compute the complexity for supersymmetric field theories for simple harmonic oscillators up to higher orders of quantum corrections in terms of the non-dynamical auxiliary field. }
	
	\item{In \underline{section~\ref{sec:8}}, we will compare the results of complexity between supersymmetric and non-supersymmetric quantum field theory for SHO and IHO in table \ref{table:a} and \ref{table:b}. }
	
	\item{In \underline{section~\ref{sec:9}}, we will provide the detailed numerical and graphical analysis of the results for both SHO and IHO using the prescription of supersymmetric Morse quantum field theory. }
	
	\item{In \underline{section~\ref{sec:10}}, we will derive the universal relation between the circuit complexity and quantum chaos expressed in terms of OTOC function for supersymmetric quantum field theory prescription using Morse function.}
	
	\item {Finally \underline{section~\ref{sec:11}}, we will conclude our result and comment on how the behaviour of complexity for the supersymmetric model is already implemented in the structure of a manifold.}

\end{itemize}

\textcolor{Sepia}{\section{\sffamily Circuit Complexity for dummies }\label{sec:2}}

The notion of circuit complexity was first introduced in information theory to find the minimum number of gates to get the desired state from an initial state. It involves acting the initial state with a unitary operator or a set of quantum gates to obtain the desired target state.

\begin{equation}
\large{\ket{\psi_{T}} = U\ket{\psi_{i}} }
\end{equation}

There exists many such set of unitary operators to get the desired final state but to find a minimum number of such operations to execute the task is what constitutes an optimal quantum circuit. Taking inspiration from this Nielsen and collaborators have developed this idea further and have taken a geometric approach to finding the most optimal quantum circuit to get the desired target state via unitary operations in physics \cite{dowling2006geometry, Jefferson:2017sdb,nielsen2005geometric, Auzzi:2020idm, Roberts:2016hpo, Choudhury:2020lja,Brown:2015bva,Yang:2017nfn,Brown:2016wib,Khan:2018rzm,Adhikari:2021pvv},

\begin{equation}
\large{U(t) = \mathcal{P}\exp{\left(-i\int_0^t H(t)dt\right)} \ \text{where} \ H(t) = \sum Y^{I}(t)M_{I}}, \label{jt}
\end{equation}
where $M_{I}$ are the Pauli matrices and $Y^{I}(t)$ are referred to as the control function that decides the nature of gate that will act at a certain value of parameter t. \textcolor{red}{This setup is a toy model for fermionic 1+1 dimensional theory, for higher dimension appropriate operators and their generators are considered. Given left hand side is the Unitary operator the right hand side will have the particular generators $M_I$ and their coefficients $Y_I$ which are tangent velocities in the Unitary operator space.} This approach of identifying the action of quantum gates via control function could be morphed in terms of finding the extremal curves, i.e. geodesics. Hence we define a cost function $F(U(t),\dot{U(t)})$ which is a function of the unitary operator $U$ and a vector at a point in the tangent space formed by unitaries. The idea is to minimize this cost function for various possible paths, which is described by the following expression:

\begin{equation}
\large{\mathcal{D}(U(t)) = \int_0^1\left( F(U(t), \dot{U(t)}\right)dt}
\end{equation}

The task now is to determine the cost function which as described above counts the number of gates to construct the optimal quantum circuit. A general class of cost functions are:

\begin{equation}
\large{F_{p}(U, Y) = \sum p_{I}|Y^{I}|,} 
\end{equation}
\begin{equation}
 \large{F_{q}(U, Y) = \sqrt{\sum q_{I}(Y^{I})^{2}},}\label{kk}
\end{equation}
where minimizing \textcolor{red}{$F_{2} = \sqrt{\sum (Y^{I}(s))^{2}}$ which is a quadratic cost functional} would give us the expression for the length of geodesic of a Riemann surface. The length of the geodesic traced out by intermidiate states in constructing target state via mininmum number of gates would give us an expression for complexity-
\begin{equation}
\large C = \underset{Y(s)}{min} \int l(ds) = \underset{Y(t)}{min}\int dt\sqrt{(\partial_{t}Y(t))^{2}}.
\end{equation} 
The gate action on the reference state could be written in terms of the generators decided by the Hamiltonian and an arbitrary small parameter $\epsilon$ as:
\begin{equation}
\large U = \exp[{M_{IJ}\epsilon}]
\end{equation}
\textcolor{red}{where $M_{IJ}$ are the suitable generators decided by the hamiltonian of the model. We can choose from a class of generators that performs the Unitary operation of taking a reference state to a target state. We can then find explict expression for the velocity $Y_{I}$ by rewritting equation \eqref{jt} in the following form}:
\begin{equation}
\large Y_{I}(s) = i(\partial_{s} U(s))U^{-1}(s),
\end{equation}
where we have used,  $Tr(M_{I}M^{J}) = \delta_{IJ}$. The general expression for the cost function could be further written as:
\begin{equation}
\large Y^I(s) = \text{Tr}[i(\partial _s U(s))(U^{-1}(s)M_I^{T}]
\end{equation}
We have written the expression for cost function in terms of the unitary operators which will explicitly demonstrate the relation between it and the Morse function for the supersymmetric Hamiltonian $H_{t}$ to compute the circuit complexity for supersymmetric field theories.

\textcolor{Sepia}{\section{\sffamily Lie algebra formulation of the Inverted Harmonic Oscillator (IHO) }\label{sec:3}}
To work out the circuit complexity in various models researchers have considered a simple, exactly solvable system known as the Inverted Harmonic Oscillator (IHO), which is described by the Hamiltonian function:
\begin{equation}
 \large{H(p,x)=\frac{p^{2}}{2} - \frac{1}{2}\omega^{2}x^{2}}.
\end{equation} 
This encapsulates the sensitivity to initial conditions exhibited by the chaotic systems \cite{Bhattacharyya:2020art,Subramanyan:2020fmx, Barton:1984ey}. The Inverted Harmonic Oscillator (IHO) differs from the Simple Harmonic Oscillator (SHO) in many ways. For instance, the energy spectrum of IHO has a continuous energy spectrum, whereas the regular SHO has a discrete energy spectrum $(n+\frac{1}{2})\hbar\omega$. While the SHO provides a good description of the deviations from the stable equilibrium, the IHO models the decay from an unstable equilibrium.
The IHO in recent times has been proven to be incredibly useful to show the equivalence among various diverse fields. The IHO appears in the quantum hall effect and in the mechanism of Rindler Hamiltonian, whose time evolution would give rise to the Hawking-Unruh effect \cite{Hawking:1974rv, Birrell:1982ix, Crispino:2007eb}. The equivalence between the two phenomena can be shown in terms of the isomorphism of the underlying lie algebra \cite{Read:2010epa, Fecko:2006zy}.

\par{In this paper, our main objective will not be to formally describe the lie algebra isomorphism but instead using that to calculate the circuit complexity over a manifold via Morse function for supersymmetric quantum field theories. To do this, first we have to show how the potential of IHO forms the generators of Lie algebra. Fortunately such construction has already been carried out and we will briefly review it below, for details readers can refer to \cite{Subramanyan:2020fmx}. We start with the settings of the quantum hall effect (QHE), the three quadratic potentials generate the Hamiltonian dynamics in the lowest Landau level (LLL):
\begin{equation}
\large P_{LLL}V_{1}P_{LLL} = \lambda(R_{x}^{2} + R_{y}^{2}), 
\end{equation}
\begin{equation}
\large P_{LLL}V_{2}P_{LLL} = \lambda(R_{X}R_{Y} + R_{Y}R_{Y}), 
\end{equation}
\begin{equation}
\large P_{LLL}V_{3}P_{LLL} = \lambda(R_{x}^{2} - R_{y}^{2}) \label{iho}.
\end{equation}
\textcolor{red}{By projecting the generators to the lowest Landau level one can find $P_{LLL}J_{ij}P_{LLL} = \frac{1}{4l^{2}_{B}}\epsilon_{ijk}\{R_{i}, R_{k} \} + \mathcal{O}(1)$. From this we see that}
\begin{equation}
\large P_{LLL}V_{1}P_{LLL} = \frac{-4\lambda l_{B}^{2}}{\hbar P_{LLL}LP_{LLL}},
\end{equation}
\begin{equation}
\large P_{LLL}V_{2}P_{LLL} = 2\lambda l_{B}^{2}P_{LLL}j_{i}P_{LLL},
\end{equation}
\begin{equation}
\large P_{LLL}V_{3}P_{LLL} = \frac{-4\lambda l_{B}^{2}}{\hbar P_{LLL}j_{j}P_{LLL}}.
\end{equation}
where $P_{LLL}$ is the projection operator to the lowest Landau level, $R_{X}, R_{Y}$ are bilinears and $j_{a}, j_{b}$ are strain generators. Therefore by projecting the potentials to the LLL both bilinears and the strain generators lead to the quadratic Hamiltonian which is similar to the electrostatic potential and IHO appears in quantum hall effect as might expected.
 
 \par{We now rename the three quadratic potentials in a quantum hall effect $V_{1}, V_{2}, V_{3}$ as $K_{1}, K_{2}, K_{3}$ and we identify:
 \begin{equation}
 \large P=\frac{R_{x}}{l_{B}},~~~~X=\frac{R_{y}}{l_{B}}.
 \end{equation}
 Consequently from \eqref{iho} , we get the Hamiltonians in the LLL to be of the following form:
 \begin{equation}
\large K_{1} = (P^{2} + X^{2}), 
\end{equation}
 \begin{equation}
\large K_{2} = (PX + XP), 
\end{equation}
 \begin{equation}
\large K_{3} = (P^{2} - X^{2}).
\end{equation}
\textcolor{red}{On the basis of LLL wave functions the potentials could be written the form of differential operators namely \cite{Subramanyan:2020fmx}}:
 \begin{equation}
\large K_{1} = \frac{1}{4}\left(-\frac{\partial^{2}}{\partial X^{2}} + X^{2}\right), 
\end{equation}
 \begin{equation}
\large K_{2} = \frac{i}{2}\left(X\frac{\partial}{\partial x} + \frac{1}{2}\right), 
\end{equation}
 \begin{equation}
\large K_{3} = \frac{1}{4}\left(-\frac{\partial^{2}}{\partial X^{2}} - X^{2}\right).
\end{equation}
 These are precisely the generators of $SL(2, R)$ Lie-algebra\cite{kerr2014notes}, which act as an area preserving deformations of a two dimensional manifold and satisfy the following non-commuting relations:
 \begin{equation}
\large [K_{2}, K_{3}] = -iK_{1}, 
\end{equation}
 \begin{equation}
\large [K_{1}, K_{2}] = iK_{3}, 
\end{equation}
 \begin{equation}
\large [K_{3}, K_{1}] = iK_{2}.
\end{equation}
 
With these techniques at hand, our main objective will now be to describe a Morse function on a manifold formed by the Lie-algebra of $SL(2, R)$, namely the potential of the IHO. Then we will show how the critical points of the Morse function on a manifold play the role of the cost function, which decides the gate in action at a particular parameter $t$ to form the optimal circuit.

 \textcolor{Sepia}{\section{\sffamily Brief review of Morse Theory }\label{sec:4}} 
In this section, we are going to give a very brief review of Morse theory. The Morse function helps classify surfaces up to homeomorphism when it passes the critical point of index $0, 1, 2.$ The index $2, 1, $ and $0$ represents the maximum, minimum and saddle points of the manifold. }

\textcolor{Sepia}{\subsection{\sffamily Definition of Morse function }\label{sec:cx}}

To define Morse function \cite{bott1982lectures,austin1995morse} we consider a manifold $\mathcal{M}$ and a function $f$ such that:
 \textcolor{Sepia}{\bf\sffamily \large \underline{Theorem~1:} } \\ \\
 \textit{ A smooth map $f: X^{n}$ $\longrightarrow$ $R$ is a Morse function if, for every critical point $p$ $\in$ $X$, $\exists$ coordinates $x_{1},...x_{n}$, and a coordinate $y$ around $f(p)$ w.r.t. which,}

 \begin{equation}
 \large f(x_{1},....,x_{n}) = -x_{1}^{2}-x_{2}^{2}-....-x_{k}^{2}+x_{k+1}^{2}+...+x_{n}^{2}.
 \end{equation}
\textit{ such that the value of the function at the critical points vanishes.} \\ \\
 \textcolor{Sepia}{\bf\sffamily \large \underline{Theorem~2:} } \\ \\ 
 \textit{ A map $f$ $\longrightarrow$ $R$ is a morse function if its critical points vanishes and the Hessian of $f$ at each critical point is non-singular}. \\ 
\par{ A Morse function on a compact manifold $X$ helps to determine it's topology, by mapping it's critical points to an axis on a one-dimensional plane,  which helps to encode a lot of information about $\mathcal{M}$ \cite{milnor2016morse}. The goal of the Morse theory is to find the invariant of the manifold by counting the critical points of chosen Morse function.}

 \textcolor{Sepia}{\subsection{\sffamily Gradient flow of Morse function }\label{sec:c}}
 
 The gradient flow of the Morse function creates a vector field on the surface of the manifold, which helps to define a notion of transport from one point to another. Let us suppose we have an integral curve $\gamma_{x}$: $R$ $\longrightarrow$ $M$ such that:
 \begin{equation}
 \large \gamma_{x}(t) = \phi_{t}(x),  ~~~~\phi_{0}(x) = 0. 
 \end{equation}
 Here $\phi$ is a smooth one-parameter group of diffeomorphism on $\mathcal{M}$. Then, we get:
 \begin{eqnarray}
 \frac{d}{dt}f(\gamma_{x}(t)) &=& \frac{d}{dt}(f~\circ~\phi_{t}(x)) \nonumber\\
               &=& df_{\phi_{t}(x)}~\circ~\frac{d}{dt}\phi_{t}(x) \nonumber\\
               &=& df_{\phi_{t}(x)}(-\nabla f)\phi_{t}(x) \nonumber\\
              &=& -||(\nabla f)\phi_{t}(x)||^{2} \le 0.
 \end{eqnarray}
 Thus the gradient flow of $f$ is decreasing down the lines of $\gamma_{x}$. One of the reasons we have defined the gradient flow of the Morse function will become apparent in the latter part of this paper, where it will play a crucial role in deriving the complexity of supersymmetric field theories for the SHO.               
 
 \textcolor{Sepia}{\section{\sffamily Circuit complexity in SUSY QFT via Morse function }\label{sec:5}}
 
To compute circuit complexity within the framework of supersymmetric quantum field theory using Morse function, we will consider $SL(2, R)$ modular curve, which encodes the potential of IHO as generators of Lie algebra and identifies it with a Riemann manifold $\mathcal{M}$, and the Morse function to the cost function as described in the previous section, and thereby associate the critical points to the action of the control function which decides, which quantum gate will be active at a particular time $t$. 

\par{The supersymmetric operators in terms of an exterior derivative and its adjoint can be described as}:
\begin{equation}
\large Q_{1} = d + d^{*}, 
\end{equation}
 \begin{equation}
\large Q_{2} = i(d - d^{*}), 
\end{equation}
\begin{equation}
\large H = dd^{*} + d^{*}d.
\end{equation}
The connection between supersymmetric quantum field theories and Derham operators could found in the ref.~\cite{Witten:1982df}. Then further by taking into account the following crucial fact: 
\begin{equation}
\large d^{2} =0= d^{*2},
\end{equation}
 we get the following subsequent supersymmetric relations in the present context, which are given by:
\begin{equation}
\large Q^{2}_{1} = Q^{2}_{2} = H,
\end{equation}
 \begin{equation}
\large Q_{1}Q_{2} + Q_{2}Q_{1} = 0.
\end{equation}

Now let us consider a Morse function $f$ on the surface of the manifold $M$, and $t$ be a real number. Then,
\begin{equation}
\large d_{t} = \exp(-ht)~d~\exp(ht), 
\end{equation}
 \begin{equation}
\large d^{*}_{t} = \exp(-ht)~d^{*}~\exp(ht).
\end{equation}
Here we can show that:
\begin{equation}
\large d^{2}_{t} =0= d^{*2}_{t},
\end{equation}
 using which we get the following expressions:
 \begin{equation}
\large Q_{1t} = d + d^{*}, 
\end{equation}
 \begin{equation}
\large Q_{2t} = i(d_{t} - d^{*}_{t}), 
\end{equation}
 \begin{equation}
\large H_{t} = d_{t}d^{*}_{t} + d^{*}_{t}d_{t}. 
\end{equation}

 We will now explicitly calculate the formula for $H_{t}$ in terms $f$, to understand how critical points come into the picture. Let, $v^{k}(p)$ and $v^{k}{*}$ be an orthonormal basis of tangent vectors and the corresponding dual vectors at each point $p$ in $\mathcal{M}$. \textcolor{red}{The $a^{k}$ acts as an interior multiplication $\psi \longleftarrow i(a^{k})\psi$ as an operator on the exterior algebra at $p$. Similarly  $a^{{k}{*}}$ acts as an exterior multiplication by the one form dual to $a^{k}$ which is the adjoint operator}. The $a^{k}$ and $a^{{k}{*}}$ could be regarded as creation and annihilation operators in the present context. We could calculate the covariant second derivative of Morse function $f$ in the dual basis of $v^{k}$ as $\displaystyle\left(\frac{D^{2}}{Dx^{i}Dx^{j}}\right)f$,
with these accords, one could then calculate the Laplacian operator $H_{t}$ acting on $p$ forms on manifold $\mathcal{M}$:
\begin{equation}
\large H_{t} = dd^{*} + d^{*}d + t^{2}(df)^{2} + \sum t\left(\frac{D^{2}f}{Dx^{i}Dx^{j}}\right)[v^{{*}{i}}, v^{j}].\label{fg}
\end{equation}
Here we define $(df)^{2}$ by the following expression:
\begin{equation}
\large (df)^{2} = g^{ij}\left(\frac{df}{dx^{i}}\right)\left(\frac{df}{dx^{j}}\right).
\end{equation}
which is the square of the gradient of the Morse function $f$, measured with respect to the Riemannian metric $g^{ij}$ of $M$. Here the term $t^{2}(df)^{2}$ plays the role of IHO potential, and also the critical points for the Morse function $f$ lies, where we have 
\begin{equation}
\large df=\sqrt{g^{ij}\left(\frac{df}{dx^{i}}\right)\left(\frac{df}{dx^{j}}\right)}=0.\label{cc}
\end{equation}
\par{To compute circuit complexity in supersymmetric quantum field theory, one could see that the gradient of the Morse function vanishes exactly at the critical points which in turn for supersymmetric field theories are the energy eigenvalues of $H_{t}$ thus the Morse function exactly picks up those eigenstates in the field space, as by supersymmteric charges $Q$ to act on an arbitary reference state, thereby from \eqref{kk} and the definition of $Y_{I}(t)$, we see that the Morse function defined on the surface of the manifold exactly describes the function of $F(U(t), \dot{U(t)})$ for supersymmetric field theories, hence we make the following identification.},
\begin{equation}
\large F(Q(t), \dot{Q(t)}) \longrightarrow f.
\end{equation}
The role of $F(Q(t), \dot{Q(t)})$ is to count the number of gates required to construct the optimal quantum circuit. Still, in the present context, it is to find the minimum of supersymmetric charges estimated by the Morse function based on the number of critical points on the surface of manifold $\mathcal{M}$ to get:
\begin{equation}
\large Q_{1}\ket{E^{n}}_{bosonic} = E^{n-i}_{bosonic}.
\end{equation}

where $E^{n-i}$ is the $(n-i)^{th}$ energy eigenstate. One could see in the argument mentioned above that eigenfunction of $H_{t}$ for large $t$, are concentrated near the critical points of $f$ therefore, for the identification, we could say that the eigenvalues of the cost function are the critical values of $SL(2, R)$ Riemann surface.

\par{\textcolor{red}{One can now see that except at the critical points $dh = 0$ the potential energy can be expressed as $V(\phi) = t^{2}dh^{2}$ for very large $t$. Therefore the eigenfunctions of $H_{t}$ for very large $t$ are concentrated near the critical points of $h$ and admits an asymptotic eigenvalue expansion in powers of $1/t$. The eigenvalues of $H_{t}$ acting on $p$ forms for large $t$ can be expressed in the following way \cite{Witten:1982im}}:}
\begin{equation}
\large \lambda^{n}_{p}(t) = t\left(A^{n}_{p} + \frac{B^{n}_{t}}{t} + \frac{C^{n}_{p}}{t^{2}} + ...\right).
\end{equation}

Thus we could see that for large $t$, the above equation, within few orders, agrees with the computations of out-of-time-order-correlations (OTOC) from supersymmetric quantum mechanics computed in \cite{Bhagat:2020pcd}. 

Now to calculate the general expression for circuit complexity, we will make use of the Hessian matrix. Remember, a function $f$ is a morse function if and only if the Hessian of $f$ at each critical point is non-singular.
This is described as:
\begin{equation}
\large H(f(x)) = j(\nabla f(x))
(Hf)_{{i}{j}} = \frac{\partial^{2} f}{\partial x_{i} \partial x_{j}}.
\end{equation}
we will now identify Hessian of $f$ in the tangent space of unitary operator as:
\begin{equation}
\large T_{p}(M) \times T_{p}(M) \longrightarrow R.
\end{equation}
and from \eqref{fg} exactly along the geodesic connnecting critical points of the Morse function, the supersymmetric hamiltonian could be written as:
\begin{equation}
H_{t} =  \left(\frac{D^{2}}{Dx^{i}Dx^{j}}\right)_{ij}f[v^{{*}{i}}, v^{j}]
\end{equation}
comparing the above to \eqref{jt} we get the desired relation between cost function and the Morse function such that
\begin{equation}
\large \frac{dU(s)}{ds} = -iY_{s}^{I} = (Hf)_{{i}{j}},
\end{equation}
where $Y_{s}^{I}$ is the control function, that decides the action of  of operators acting on the reference state to make the optimal circuit. 

Consequently, we compute:
\begin{equation}
\large {\cal D}(Q)(s) = \sqrt{(Y^{11})^{2} + (Y^{12})^{2} + (Y^{21})^{2} + (Y^{22})^{2}},
\end{equation}
and the corresponding circuit complexity can be computed as:
\begin{equation}
\large \boxed{\boxed{\mathcal{C}(Q) = |H(f(x))|= |j(\nabla f(x))
(Hf)_{{i}{j}}|=\left|\frac{\partial^{2} f}{\partial x_{i} \partial x_{j}}\right|}}~.
\end{equation}
\textcolor{Sepia}{\section{\sffamily Effect on circuit complexity from IHO perturbation theory}\label{sec:6}}

To explicitly calculate circuit complexity for IHO in a supersymmetric case, we will use the concept of the Witten index and show how the critical point near which the eigenvalues of $H(t)$ are concentrated grows exponentially with the superpotential. 

\par{The excited states of supersymmetric quantum field theory always come with pair of states, this could be seen in the algebra of supersymmetry which has no one-dimensional representation \cite{Drees:1996ca}:}
\begin{equation}
\large \{Q, Q^{*}\} = 2H \ \ ~~such ~that~~ \ \ Q^{2} = Q^{*2} = 0~.
\end{equation}
The Witten index $\displaystyle {\rm Tr}[(-1)^{F}e^{-\beta H}]$, where $(-1)^F$ is the well known fermion number operator, carries interesting information about the ground state of supersymmetric quantum system especially when it is non-zero, i.e. the system has at least one ground state when the Witten index is non-zero, however, it doesn't make any comment on the number of ground states of the system when it is zero \cite{Witten:1981nf}. We will take the Hamiltonian of the supersymmetric quantum system in IHO regime by replacing $W(x)$ by $iW(x)$ and show how the critical points of Morse function or complexity grow exponentially with respect to the superpotential, which is given by: 
\begin{equation}
\large H(P,W) = \frac{P^{2}}{2} - \frac{W^{2}}{2}~.
\end{equation}
have a ground state wave function, which is defined as:
\begin{equation}
\large \psi = \exp\left(-i\sigma_{2}\int_\infty^{\infty} Wdx\right)~.
\end{equation}
Now by doing supersymmetry on the two dimensional manifold the Witten index can be expressed by the following expression \cite{Witten:1982im}:
\begin{equation}
\large {\rm Tr}[(-1)^{F}\exp(-\beta H)] = \chi(M)~.
\end{equation}
where $\chi(M)$ is the Euler characteristic of the manifold, thereby using the strong Morse inequality, we further get:
\begin{equation}
\large \chi(M)={\rm Tr}[(-1)^{F}\exp(-\beta H)] =\sum (-1)^{F}\exp(-\beta H)= \sum (-1)^{\gamma}C^{\gamma}~.
\end{equation}
where $C^{\gamma}$ is the number of critical points of index $\gamma$, using which one can find out the following simplified expression:
\begin{equation}
\large C^{\gamma}=\left|(-1)^{F-\gamma}\exp{\left(-\beta \left(\frac{P^{2}}{2} - \frac{W^{2}}{2}\right)\right)}\right|~.
\end{equation}
 Circuit complexity for higher-order interacting term could be calculated in terms of super-field formalism $\Phi$, in which all super-partners related by SUSY transformations could be treated as a single field \cite{Weinberg:2000cr,Gates:1983nr,Quevedo:2010ui}. Scalars and fermions related by supersymmetry correspond to different components of super-field. The most general form of super-fields in terms of super-space variables can be expressed as follows:
 \begin{equation}
\large \Phi(z) = \phi(x) + \theta\psi + \bar{\theta}\bar{\chi}(x)+ \theta\sigma^{\mu}\bar{\theta}A_{\mu}(x) + \theta^{2}F(x)~.
\end{equation}
Now, to write the super-field in terms of any single field component, we will apply the SUSY transformations by applying the operator $\displaystyle i(\xi Q + \bar{\xi Q})$ under the projection:
 \begin{equation}
 \large \theta = \bar{\theta} = 0,
 \end{equation}
Expressing the result in terms of other components:
 \begin{equation}
 \large \delta_{\xi}\phi(x) = i(\xi D + \bar{\xi}\bar{D})\Phi|_{\theta = \bar{\theta} = 0} = -\xi \psi(x) - \bar{\xi} \bar{\chi(x)},
 \end{equation}
where $D$ is a super-covariant derivative of the super-field which anti-commutes with the supersymmetric charges $Q$ and $\bar{Q}$ and under transformations maps super-field to super-field, given by:
 \begin{equation}
 \large \delta_{\xi}D_{\alpha}\Phi(z) = D_{\alpha}\delta_{\xi}\Phi(z) = i(\xi Q + \bar{\xi}\bar{ Q})D_{\alpha}\Phi(z)
 \end{equation}
They satisfy the same algebra as supersymmetric charges, thus the scalar component $\phi(x)$, the super-field can be recovered by exponentiating supersymmetry transformations with $\theta$ as the parameter:
\begin{equation}
 \large \Phi = \exp{(-\delta_{\theta})}\phi(x),
 \end{equation}
Thus, the super-fields can be constructed by applying the operator $\displaystyle \exp{(-\delta_{\theta})}$ to any component field. Hence under the previously mentioned projection $\displaystyle \theta = \bar{\theta} = 0$ the complexity for higher order interactions in the super-potential could be written is as follow: \\
 
\begin{equation}
\boxed{\boxed{\large \textcolor{red}{\bf Complexity~ for~ \phi^{3}~ term:}~C^{\gamma}=(-1)^{F-\gamma}\exp{\left(-\beta \left(\frac{P^{2}}{2} - \frac{m}{2}\phi^{2}- \frac{\lambda}{3}\phi^{3}\right)\right)}}}~
\end{equation}

\begin{equation}
\boxed{\boxed{\large \textcolor{red}{\bf Complexity~ for~ \phi^{2} + \phi^{4}~ term:}~\large C^{\gamma}=(-1)^{F-\gamma}\exp{\left(-\beta \left(\frac{P^{2}}{2} - \frac{m}{2}\phi^{2}- \frac{\lambda}{4}\phi^{4}\right)\right)}}}~
\end{equation}

\textcolor{Sepia}{\section{\sffamily Effect on circuit complexity from SHO perturbation theory}\label{sec:7}}

To calculate the circuit complexity for supersymmetric quantum field theory, it is convenient to work in the super-space formalism \cite{Shirman:2009mt, Wess:1974jb}, i.e. we extend the $4$ commuting space-time coordinates ${x_{\nu}}$ to $4$ commuting and $4$ ant-commuting coordinates $\displaystyle {x_{\nu}, \theta^{\alpha}, \bar{\theta^{\alpha}}}$. These new coordinates satisfy the following anti-commuting relations:
\begin{equation}
 \large\{\theta_{\alpha}, \theta_{\dot{\beta}}\} = \{\theta_{\alpha}, \theta_{\beta}\} = \{\theta_{\dot{\alpha}}, \theta_{\dot{\beta}}\} = 0.
 \end{equation}
Now, any super-multiplet in super-space coordinates could be communicated in terms of super-fields and can be expressed as:
\begin{equation}
 \large\Phi = \phi(x) - i\theta \sigma^{\mu} \bar{\theta}\partial_{\mu}\phi(x) - \frac{1}{4}\theta^{2}\bar{\theta^{2}}\partial^{2}\phi(x) + \sqrt{2}\theta \eta + \frac{i}{\sqrt{2}}\theta^{2}\partial_{\mu}\eta \sigma^{\mu}\bar{\theta} + \sqrt{2}\theta^{2}F(x).
 \end{equation}
where $\eta$ is a Weyl fermion having $4$ off-shell degrees of freedom and $\sigma$ are the Pauli matrices and $\phi$ is a complex scalar having two degrees of freedom. The supersymmetric Lagrangian remains invariant even after the addition of the term, i.e.
\begin{equation}
 \large\delta \mathcal{L} = \mu^{2}F + h.c.
 \end{equation}
The $F$ term is an auxiliary complex bosonic field with two off-shell degrees of freedom to match the four off-shell degrees of freedom of a Weyl fermion. Also, the $F$ term is an order parameter for SUSY breaking, substituting $F$ = -$\mu^{2}$ the ground state will not be invariant, and supersymmetry will be spontaneously broken.

The Lagrangian in terms of components fields up to second order is:
\begin{equation}
 \large\mathcal{L} = |\partial_{\mu}\phi|^{2} + i\eta^{\dagger}\partial_{\mu}\bar{\sigma^{\mu}}\eta + |F|^{2} + \left(m F \phi - \frac{m}{2}\eta \eta + h.c\right).
 \end{equation}
We will now identify the auxiliary field $F$ with the gradient of the Morse function which passes through every critical point on the surface, such that these points act as an order parameter for SUSY breaking. By doing the coordinate transformation, $U(u, v)$ = $(x(u, v), \theta(u, v))$ we could define:
\begin{equation}
 \large h(u, v) = g \circ U(u, v)
 \end{equation} 
 such that: 
 \begin{equation}
 \large \frac{\partial}{\partial x} g = F.
 \end{equation} 
 The eigenvalues of $H(t)$ are concentrated at the critical points of $g$, hence precisely at the critical point where:
 \begin{equation}
 \large dg = F=0
 \end{equation} 
represents the scalar potential of the theory with no supersymmetric ground state. Then Hessian of $h$ can be calculated as:
\begin{equation}
 \large \frac{\partial^{2} h}{\partial u^{2}} = F x_{u}^{2} + 2F x_{u}\theta_{u},
 \end{equation} 
 \begin{equation}
 \large \frac{\partial^{2}h}{\partial v^{2}} = F x_{v}^{2} + 2F x_{v}\theta_{v},
 \end{equation} 
 \begin{equation}
 \large \frac{\partial^{2}h}{\partial u\partial v} = F x_{u}x_{v} + F x_{v}x_{u} + F\theta_{u}\theta_{v} + F x_{uv} + F x_{v}\theta_{u}.
 \end{equation} 
the above dependence on the derivative of super-space variable with respect to $u$ and $v$ is actually the Jacobian due to the change of the variables as mentioned above. Now solving the equation of motion for $F$, the circuit complexity can be evaluated as:
\begin{equation}
 \large \boxed{\boxed{\large \mathcal{C(Q)} ={\cal F}_1(x_u,x_v,\theta_u,\theta_v,\phi)- {\cal I}_1(x_u,x_v,x_{uv},\theta_u,\phi) }},
\end{equation}
where the newly introduced function $\displaystyle {\cal F}_1(x_u,x_v,\theta_u,\theta_v,\phi)$ and $\displaystyle {\cal I}_1(x_u,x_v,x_{uv},\theta_u,\phi)$ are defined by the following expressions:
\begin{equation}
\large {\cal F}_1(x_u,x_v,\theta_u,\theta_v,\phi):= \left(\frac{m\phi}{2}x_{u}^{2} + 2\frac{m\phi}{2} x_{u}\theta_{u}\right)\left(\frac{m\phi}{2} x_{v}^{2} + 2 \frac{m\phi}{2}x_{v}\theta_{v}\right),
\end{equation}
\begin{equation}
\large {\cal I}_1(x_u,x_v,x_{uv},\theta_u,\phi):=\left( \frac{m\phi}{2}x_{u}x_{v} + \frac{m\phi}{2}x_{v}x_{u} + \frac{m\phi}{2} x_{uv} + \frac{m\phi}{2}x_{v}\theta_{u}\right)^{2}.
\end{equation}

\textcolor{Sepia}{\subsection{\sffamily Circuit complexity for $\phi^{3}$ term }\label{sec:g}}

To calculate the complexity for higher order interacting terms we will follow the same procedure as above, notice that circuit complexity for SUSY field theories only depends on the value of auxiliary field which also act an order parameter for soft SUSY breaking. The Lagrangian for cubic interactions could be defined as:
\begin{equation}
\large \mathcal{L} = |\partial_{\mu}\phi|^{2} + i\eta^{\dagger}\partial_{\mu}\bar{\sigma^{\mu}}\eta + |F|^{2} + \left(m F \phi + \lambda F\phi^{2} - \frac{m}{2}\eta \eta - \lambda \phi \eta \eta + h.c\right)
\end{equation}
The kinetic term is a Kh$\ddot{a}$ler potential corresponding to the $\theta\bar{\theta}$ term which is invariant under the SUSY transformations. A general Kh$\ddot{a}$ler potential could give rise to complicated terms in the Lagrangian, but for simplicity we will consider the most canonical kinetic terms. $\lambda$ here is the coupling constant, and the corresponding $F$ term is given by the following expression:
\begin{equation}
\large F(x) = - \frac{\phi(m + 2\lambda)}{2}.
\end{equation}
As described above we will now identify gradient of the Morse function with absolute value of the auxiliary field $F$ which is invariant under the SUSY transformations, and then compute the hessian of the Morse function and thereby the complexity as described above:
\begin{equation}
 \large \boxed{\boxed{\large \mathcal{C(Q)} ={\cal F}_2(x_u,x_v,\theta_u,\theta_v,\phi)- {\cal I}_2(x_u,x_v,x_{uv},\theta_u,\phi) }},
\end{equation}
where the newly introduced function $\displaystyle {\cal F}_2(x_u,x_v,\theta_u,\theta_v,\phi)$ and $\displaystyle {\cal I}_2(x_u,x_v,x_{uv},\theta_u,\phi)$ are defined by the following expressions:
\begin{equation}
{\cal F}_2(x_u,x_v,\theta_u,\theta_v,\phi):= \left(\frac{\phi(m + 2\lambda)}{2}x_{u}^{2} + 2\frac{\phi(m + 2\lambda)}{2} x_{u}\theta_{u}\right)\left(\frac{\phi(m + 2\lambda)}{2} x_{v}^{2} + 2 \frac{\phi(m + 2\lambda)}{2}x_{v}\theta_{v}\right),
\end{equation}
\begin{equation}
 {\cal I}_2(x_u,x_v,x_{uv},\theta_u,\phi):=\left( \frac{\phi(m + 2\lambda)}{2}x_{u}x_{v} + \frac{\phi(m + 2\lambda)}{2}x_{v}x_{u} + \frac{\phi(m + 2\lambda)}{2} x_{uv} + \frac{\phi(m + 2\lambda)}{2}x_{v}\theta_{u}\right)^{2}.
\end{equation}

\textcolor{Sepia}{\subsection{\sffamily Circuit complexity for $\phi^{4}$ term }\label{sec:y}}
Super-potential allows to introduce a variety of supersymmetric interactions, here we will study complexity for quartic interaction terms namely $\phi^{2} + \phi^{4}$. The representative Lagrangian involving quartic interaction in terms of components of super-fields is given by:
\begin{equation}
\large \mathcal{L} = |\partial_{\mu}\phi|^{2} + i\eta^{\dagger}\partial_{\mu}\bar{\sigma^{\mu}}\eta + |F|^{2} + \left(m F \phi + \lambda \phi^{3} - \frac{m}{2}\eta \eta - \lambda \phi^{2}\eta\eta + h.c\right).
\end{equation}
Here $F$ is again representing the auxiliary field as described above, responsible for soft SUSY breaking. By solving the equation of motion for $F$ term we get:
\begin{equation}
\large F = -\frac{m\phi + \lambda \phi^{3}}{2}.
\end{equation}

 Again identifying the auxiliary field by gradient of the Morse function as described in previous section, we could compute the circuit complexity by computing the Hessian of the Morse function for the corresponding perturbed term which then gives:
 \begin{equation}
 \large \boxed{\boxed{\large \mathcal{C(Q)} ={\cal F}_3(x_u,x_v,\theta_u,\theta_v,\phi)- {\cal I}_3(x_u,x_v,x_{uv},\theta_u,\phi) }},
\end{equation}
where the newly introduced function $\displaystyle {\cal F}_3(x_u,x_v,\theta_u,\theta_v,\phi)$ and $\displaystyle {\cal I}_3(x_u,x_v,x_{uv},\theta_u,\phi)$ are defined by the following expressions:
\begin{equation}
{\cal F}_3(x_u,x_v,\theta_u,\theta_v,\phi):= \left(\frac{m\phi + \lambda \phi^{3}}{2}x_{u}^{2} + 2\frac{m\phi + \lambda \phi^{3}}{2} x_{u}\theta_{u}\right)\left(\frac{m\phi + \lambda \phi^{3}}{2} x_{v}^{2} + 2 \frac{m\phi + \lambda \phi^{3}}{2}{v}\theta_{v}\right),
\end{equation}
\begin{equation}
 {\cal I}_3(x_u,x_v,x_{uv},\theta_u,\phi):=\left( \frac{m\phi + \lambda \phi^{3}}{2}x_{u}x_{v} + \frac{m\phi + \lambda \phi^{3}}{2}x_{v}x_{u} + \frac{m\phi + \lambda \phi^{3}}{2} x_{uv} + \frac{m\phi + \lambda \phi^{3}}{2}x_{v}\theta_{u}\right)^{2}.
\end{equation}

Hence we see that circuit complexity for supersymmetric field theories only depends on the absolute value of the auxiliary field coming from the linear term in the superpotential. The above method for calculating the complexity of SHO in a similar manner to what we have used for IHO. The number of zero eigenvalues of supersymmetric $H$ is precisely equal to the Euler number of the manifold, therefore by identifying the $F$ the term with the gradient of the Morse function, the critical points are exactly where the $F$ term becomes zero and determines the ground state of the system. 

\textcolor{Sepia}{\section{\sffamily Comparative Analysis}\label{sec:8}}
\textcolor{Sepia}{\subsection{\sffamily SHO }\label{sec:tabSHO}}
\begin{table}[!h]
\begin{center}
\footnotesize
\begin{tabular}{|p{2cm}|p{3cm}|p{4cm}|p{4cm}|p{3cm}|} 
\hline
\multicolumn{1}{|c|}{Parameter}&\multicolumn{2}{|c|}{Non-SUSY QFT}&\multicolumn{2}{|c|}{SUSY QFT}\\
\hline
-& $\phi^{4} + \phi^{2}$ & $\phi^{2}$ &$\phi^{4} + \phi^{2}$ & $\phi^{2}$ \\
\hline
Mass & Complexity for Non-supersymmetric field theories has polynomial as well as logarithmic dependence on the mass parameter as $log(m\delta)$ & Circuit complexity for Non-SUSY QFT has logarithmic support on the square of the mass parameter $log(m^{2}\delta)$, and in the infrared (IR) region it takes the form $-log^{k}(mk)$. & Complexity for Supersymmetric field theories has only polynomial dependence, namely the quadratic exponent of the mass parameter. For \textbf{large masses} we observe a decrease in rate and reaches a saturation value. & Complexity for Supersymmetric field theories in case of quadratic perturbations also has polynomial, namely the quadratic dependence on the mass parameter. \\
\hline
Topological dependence & Complexity for Non-supersymmetric field theories has a fractional reliance on the volume of lattice for interacting terms, such as $V^{\frac{1}{2}}$ for dimension $d = 3$ & Complexity due to just quadratic perturbations in Non-SUSY QFT doesn't have any topological dependence but depends on the dimension of lattice used for computations & Complexity for supersymmetric field theories due to quadratic interactions depends on the Hessian of the Morse function whose gradient has been identified with the auxiliary field and also on the critical points of the manifold and hence has topological dependence & SUSY complexity due to quadratic term in the superpotential also depends on topological parameters (Hessian of the Morse function) defined on the manifold. \\
\hline 
Dependence on the field & Complexity for Non-supersymmetric field theories depends on various parameters of a quantized field in theories and the strength of interaction with each other and normal frequency modes & Due to quadratic perturbations, complexity for Non-SUSY field theories on a lattice depends on components of the momentum vectors and the number of oscillators in the lattice formalism. It doesn't depend on the coupling parameter. & Complexity for SUSY QFT only depends on the absolute value of the non-dynamical auxiliary field, namely the $F-term$ as $F = \frac{m\phi + \lambda \phi^{3}}{2} $ which is identified as the gradient of the Morse function, which passes through every critical point on the surface & In case of quadratic perturbation complexity changes due to the shift of $F$-term which by solving the equation of motion is given by $\frac{m\phi}{2}$ and doesn't have any dependence on coupling constant \\
\hline
Growth of complexity & Complexity for perturbating term $\phi^{4}$ for dimension $d > 0$ breaks down and in the limit $\lambda \rightarrow 0$ the circuit complexity have a continuous limit & In the infrared scale i.e $\omega_{0} \ll \frac{1}{\delta}$ their is an additional logarithmic factor in the complexity and lead to divergences in the limit $\delta \rightarrow 0$. & When we change coupling $\lambda$ from 1 to -1, we observe that complexity first rises and then have a sudden dip. As we go to more negative values of lambda, we observe the rate of saturation to be faster. & In supersymmetric field theories, we have strangely observed that complexity first rapidly grows and then saturates doesn't change much when we change the coupling $\lambda$ to -1 \\
\hline
\end{tabular}
\end{center}
\caption{Comparison in circuit complexity between SUSY \& NON-SUSY QFT	for SHO}\label{table:a}
\end{table}

\newgeometry{top=2cm,bottom=3.81cm}
\begin{table}[!htb]
\begin{center}\footnotesize
\begin{tabular}[!htb]{|p{1.8cm}|p{14.65cm}|}
\hline
Parameters of SHO & Features of graph and Lyapunov \\
\hline 
General features & The graphs rise fast initially and then slowly saturate. The rate of saturation and complexity at the saturation point depends on the order for perturbation, with $\phi^4$ term being the slowest to saturate. Hence the slope is significant, and the Lyapunov exponent is expected to be larger for $\phi^4$ theory. The $\phi^3$ theory saturates quickly, giving a smaller slope than the rest and a smaller Lyapunov exponent. \\ 
\hline
Mass & For \textbf{large masses} we observe a decrease in rate and saturation value of $\phi^4$ theory. It becomes in-differentiable to the $\phi^2$ term as we approach massive fields (hence a smaller Lyapunov exponent). For \textbf{smaller masses} the $\phi^3$ graph approaches $\phi^2$, decreasing slightly in rate. We can expect a slight increment in the Lyapunov exponent.\\ 
\hline
$\lambda$ & When we make \textbf{$\lambda$ negative} we observe the saturation is slower in $\phi^2$ and $\phi^3$ perturbations. In the case of $\phi^4$ perturbation, we encounter a zero, thereby right-shifting the point of initial rise, increasing the value of the Lyapunov exponent. As we go to more negative values of lambda, we observe the rate of saturation to be faster, and we expect the Lyapunov exponent to be smaller. \\
\hline
\end{tabular}
\end{center}
\caption{Discussions on Lyapunov exponent for SHO complexity}
\end{table}

\textcolor{Sepia}{\subsection{\sffamily IHO }\label{sec:tabIHO}}
\begin{table}[!h]
\footnotesize\begin{center}
\begin{tabular}{|p{2cm}|p{7cm}|p{7cm}|} 
\hline
Parameter & Non-supersymmetric QFT ($\phi^{2}$) & Supersymmetric QFT ($\phi^{2}$) \\
\hline
Mass & Circuit complexity for Non-SUSY QFT in the regime of inverted harmonic oscillators has a quadratic dependence on a mass parameter in the exponential type function, namely the inverse cosine hyperbolic function. It has also been observed that complexity starts to increase before the critical value $\lambda = m^{2}$. & Circuit complexity for supersymmetric field theories for the inverted harmonic oscillator also has a quadratic dependence on the mass parameter in the exponential function. We also observe that as mass increases, the rate of change of complexity increases. \\
\hline
Topological dependence & Circuit complexity of Non-supersymmetric quantum field theories doesn't have any topological dependence. However, it depends on the number of oscillators and dimension of the lattice & Circuit complexity for supersymmetric field theories depends on the critical points of the manifold, and for even values of $F - \gamma$, we see that complexity increases exponentially. \\
\hline
$\lambda$ &The complexity starts to increase for $\lambda < \lambda_{c}$. At the critical point, the complexity sharply increases. Beyond the critical value $\lambda = m^{2}$, the model becomes unstable. We expect the complexity to grow rapidly with decreasing pick up time & The change in the value of $\lambda$ contributes to the rate with increasing values resulting in faster rates of increase and hence higher slopes. For negative values of $\lambda$ we observe that complexity decreases exponentially, and the model becomes irrelevant. \\
\hline 
Growth in complexity & For inverted harmonic oscillator, complexity for non-supersymmetric field theories for the initial time is nearly zero, after which it exhibits linear growth. & On the contrary, the inverted harmonic oscillator complexity doesn't exhibit any exponential or linear growth, as seen in figure (9.5). \\
\hline
\end{tabular}
\end{center}
\caption{Comparison in complexity between NON-SUSY \& SUSY QFT for IHO}\label{table:b}
\end{table}

\restoregeometry

\begin{table}
\begin{center}\footnotesize
\begin{tabular}[!htb]{|c|p{13cm}|}
\hline
Parameters of IHO & Features of graph and Lyapunov \\
\hline 
General features & We do not observe any saturation behaviour, and the complexity values rise quickly to very high values. Although we can not associate a Lyapunov exponent, we make general statements about the increase rate (hence the slope) of the graphs. We observe increased gradients upon adding perturbation terms. \\ 
\hline
$(F-\gamma)$ & For even values, the graphs are increasing exponentially, whereas, for odd values, we see negative complexity values and hence have not included them in our graphical analysis.\\ 
\hline
Momentum $p$ & The constant momentum factor acts as a scale multiplying the overall complexity value and is hence redundant. \\
\hline
Mass & For $\phi^3$ perturbation, we observe as mass increases, the rate also increases, resulting in larger slope values. This is also true in the case of $\phi^4$ theory, with the only difference being the symmetry along the vertical-axis.\\
\hline
Temperature & The dependence on temperature can be evaluated by varying $\beta$ (the inverse temperature). By varying this, we can observe that for high temperatures, the rate of increase is much lesser as compared to lower temperatures in the case of IHO. This is true for both $\phi^3$ and $\phi^4$ perturbations. One can note that this behaviour contradicts the upper bound that we can set for Lyapunov exponents giving us more incentive not to associate the slope of IHO with the Lyapunov exponent. \\
\hline
$\lambda$ & For negative $\lambda$ value, we observe a mirror inversion of the graph about vertical-axis and hence the complexity are exponentially decreasing. One can interpret the opposite behaviour of the graphs with mass and temperature variation in negative $\lambda$. The change in the value of $\lambda$ contributes to the rate with increasing values resulting in faster rates of increase and hence, higher slopes. \\
\hline
\end{tabular}
\end{center}
\caption{Discussions on Lyapunov exponent for IHO complexity}\label{tabel:d}
\end{table}

\textcolor{Sepia}{\section{\sffamily Graphical Analysis}\label{sec:9}}

With the computed formulae for complexity for IHO and SHO in the previous sections, we plot the graphs using different values of parameters.

\textcolor{Sepia}{\subsection{\sffamily SHO }\label{sec:graphSHO}}

For SHO, the parameters involved are the mass $m$ and coupling coefficient $\lambda$ for higher-order theories. We have chosen a simple linear coordinate transformation between the field variables $x$ and $\theta$ and $u$,$v$. This ensures that the Hessian is a constant, and our computations become much more straightforward. The particular coefficients of the transformation have been chosen so that complexity is positive and rising for all cases. 

\begin{figure}[!htb]
	\centering
	\includegraphics[width=17cm,height=9cm]{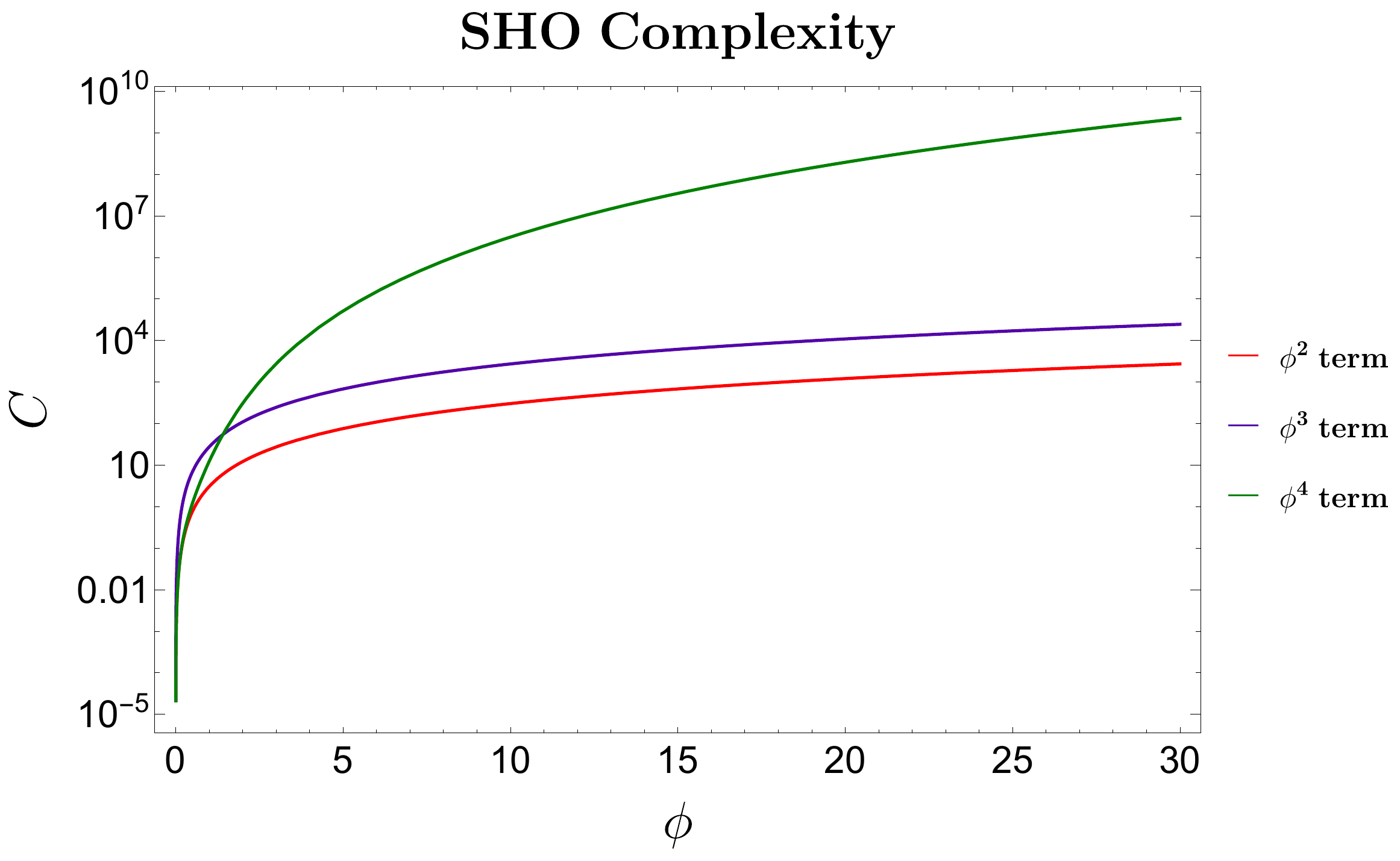}	
	\caption{Complexity against $\phi$ for $m=1$, $\lambda=1$}\label{fig:SHOm}
\end{figure}
\begin{figure}[!htb]
	\centering
	\includegraphics[width=17cm,height=9cm]{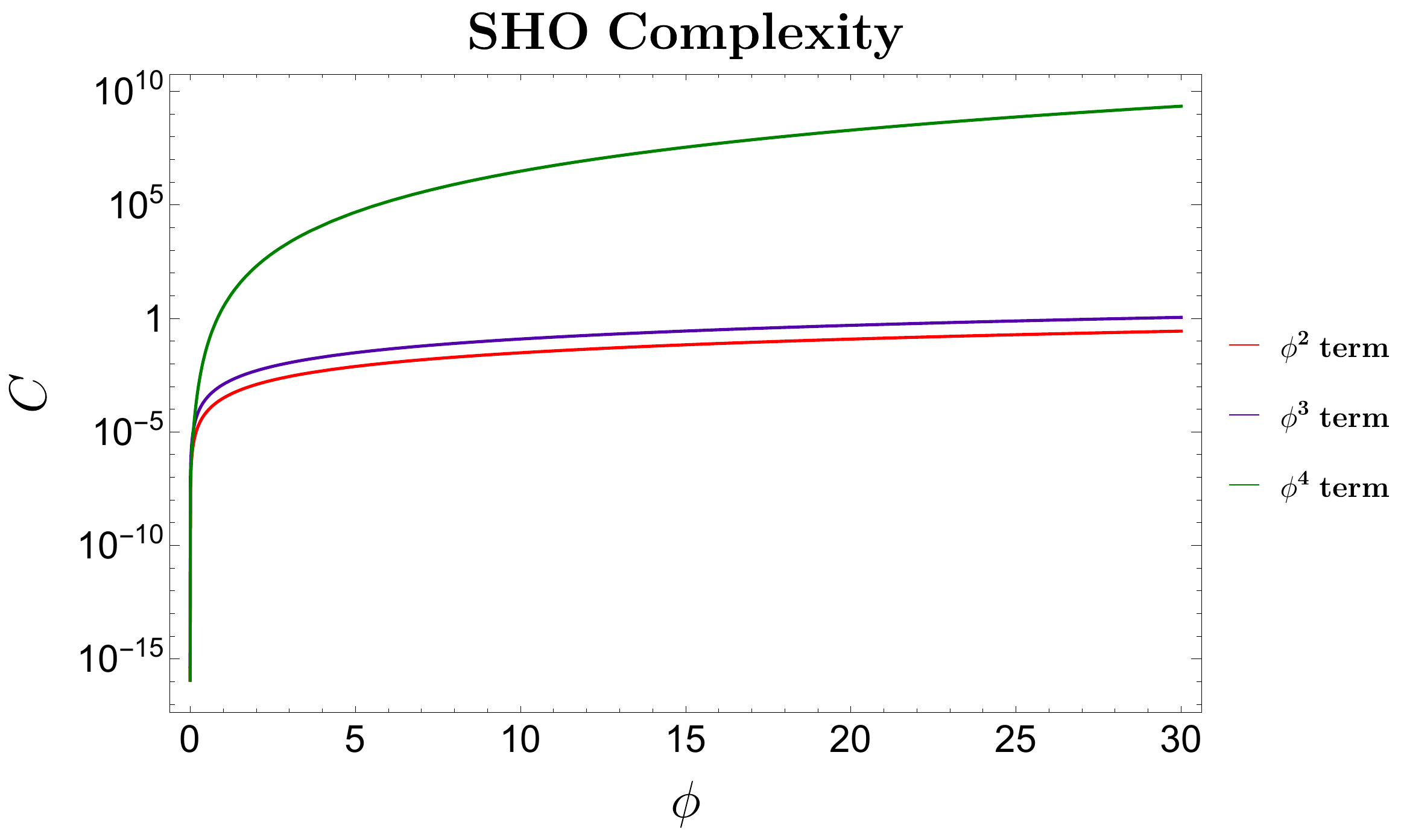}	
	\caption{Complexity against $\phi$ for $m=0.01$, $\lambda=1$}\label{fig:SHOsm}
\end{figure}
\begin{figure}[!htb]
	\centering
	\includegraphics[width=17cm,height=9cm]{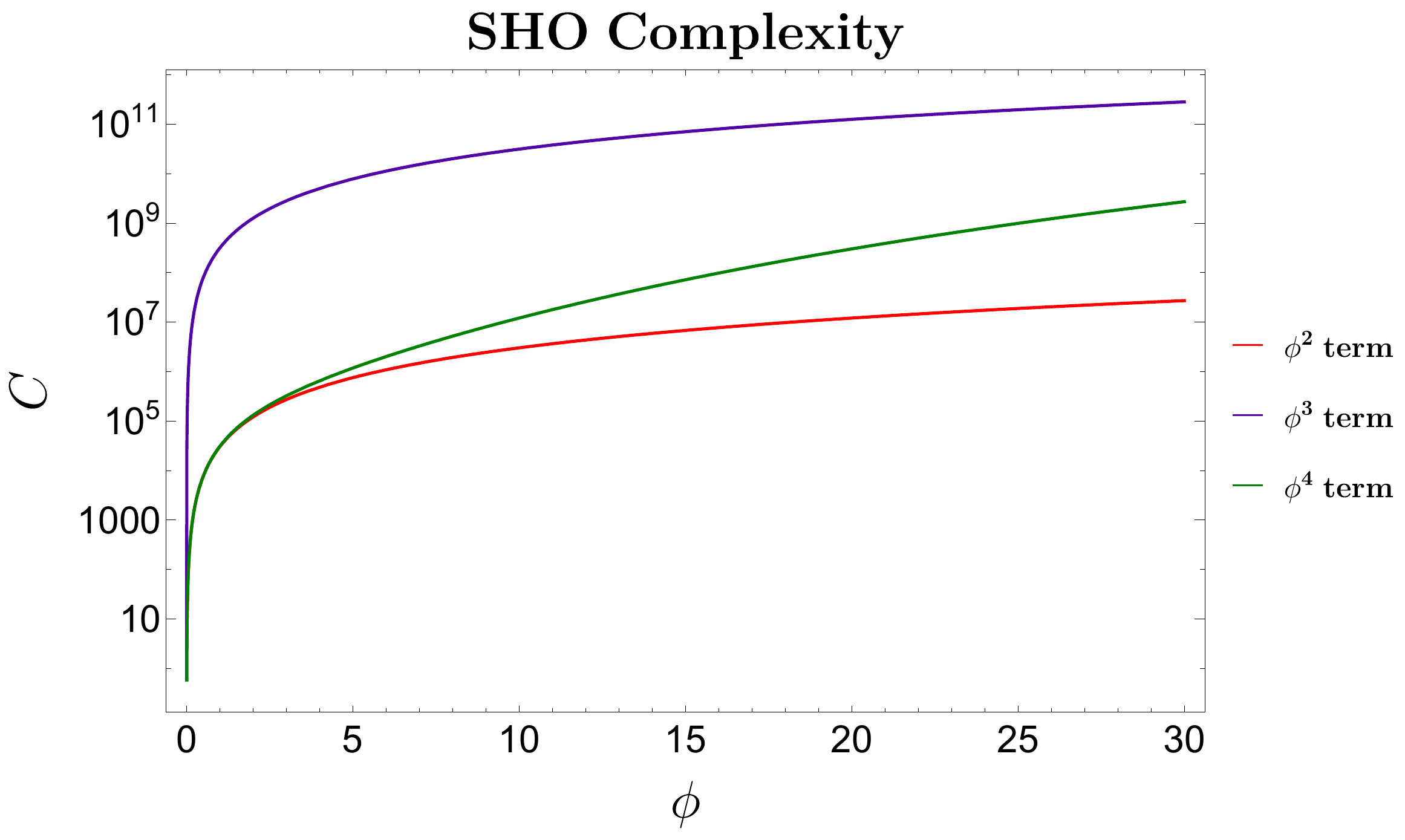}	
	\caption{Complexity against $\phi$ for $m=100$, $\lambda=1$}\label{fig:SHOlm}
\end{figure}

In \Cref{fig:SHOm} we have plotted the three different complexities for an intermediate value of mass, keeping the coupling coefficient fixed. We observe that the complexity rises and saturates as expected.

It is important to know how the complexity value behaves for different masses. For lighter particles, as we see in 
\Cref{fig:SHOsm}, we see a slight overall decrease in the complexity values, but more importantly, we observe that the graph of $\phi^3$ term tends very close to the $\phi^2$ graph. We conclude that as the mass grows smaller, the $\phi^3$ graph will inch closer and be in-differentiable to the $\phi^2$ graph. 

For larger masses, we see a stark difference. Here the graph of the $\phi^4$ theory inches closer to the $\phi^2$ graph as seen in \Cref{fig:SHOlm} and we can once again expect that it will become in-differentiable for much larger masses.

\begin{figure}[!htb]
	\centering
	\includegraphics[width=17cm,height=9cm]{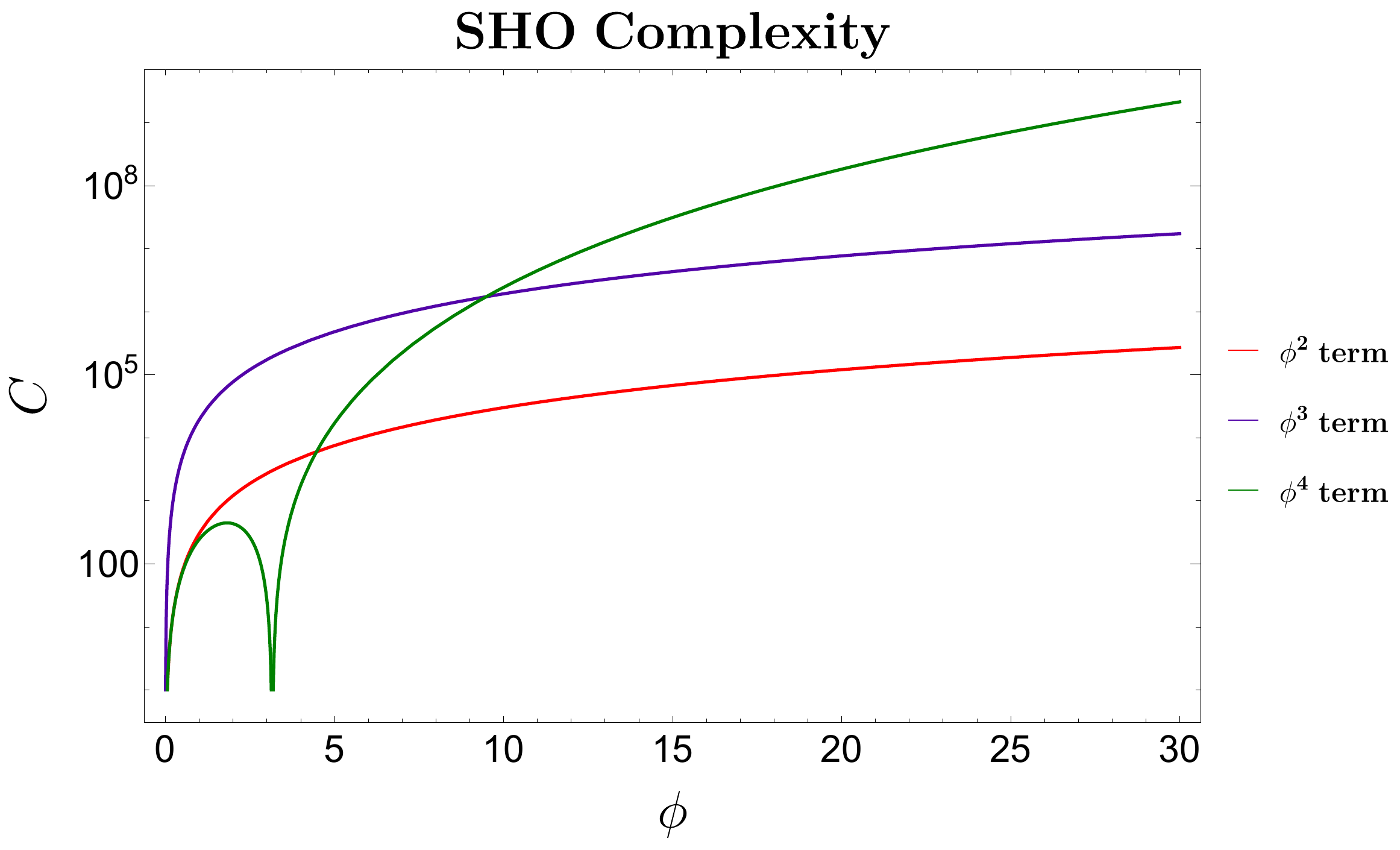}	
	\caption{Complexity against $\phi$ for $m=1$, $\lambda=-1$}\label{fig:SHOneg}
\end{figure}

The other important parameter that we need to vary is the value of $\lambda$, and we start by seeing what will happen when it is made negative as seen in \Cref{fig:SHOneg}. We know the behaviour of the complexity remains the same, i.e. it grows and saturates as we go right. An exciting behaviour occurs in the $\phi^4$ graph - We see an initial rise and dip before the rise and saturation. This indicates that we have a zero in the $\phi^4$ theory.

We observe no behavioural changes when we vary the specific value of $\lambda$ apart from the fact that for higher values, it saturates much quickly (pointing to a larger Lyapunov exponent) and saturates slower for smaller values (smaller Lyapunov exponent).

\textcolor{Sepia}{\subsection{\sffamily IHO}\label{sec:graphIHO}}

For the plots of IHO, we need to keep in mind that there are many more parametric values. We will do a graphical analysis mainly by varying mass and temperature. We will merely state the redundancy involved in other parameters and therefore explain our choice of fixing it. 

\begin{itemize}

\item If one looks closer to the formulae given for IHO in \Cref{sec:6} we see the value of the momentum - $p$ just adds an overall factor that multiplies the function, thereby acting as a scaling factor. Hence we can set this to $p=1$ without any issues.
\item Another important parameter that might affect our plots is whether the value of $F-\gamma$ is Odd or Even. When it is even, we have positive complexity, and when it is odd, we find that we are dealing with negative complexity values, which we safely ignore in the present context of the discussion. 

\end{itemize}

\begin{figure}[!htb]
	\centering
	\includegraphics[width=17cm,height=9cm]{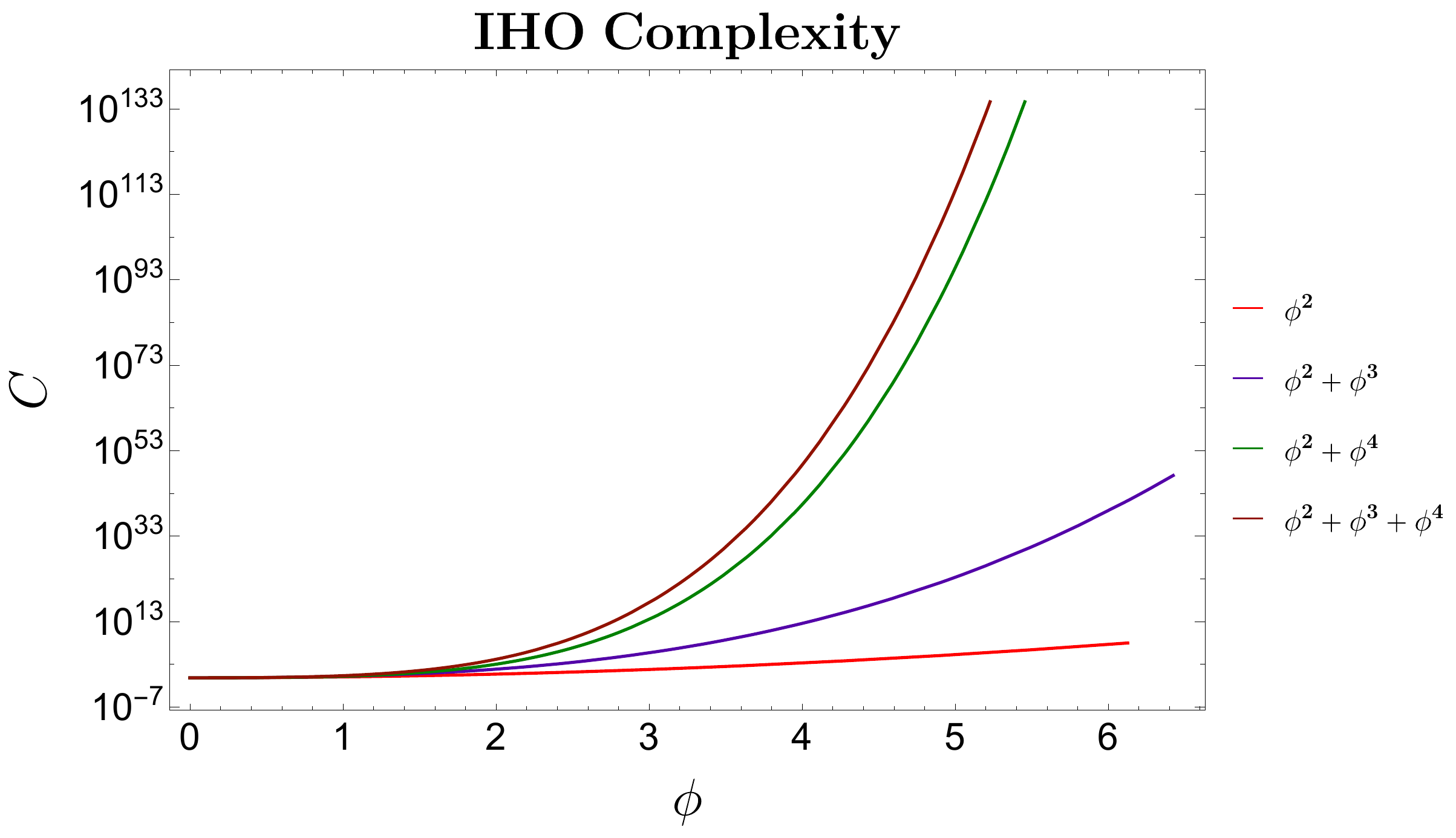}	
	\caption{Complexity against $\phi$ for different perturbations of field with $\lambda=1$}\label{fig:IHO}
\end{figure}
\begin{figure}[!htb]
	\centering
	\includegraphics[width=17cm,height=9cm]{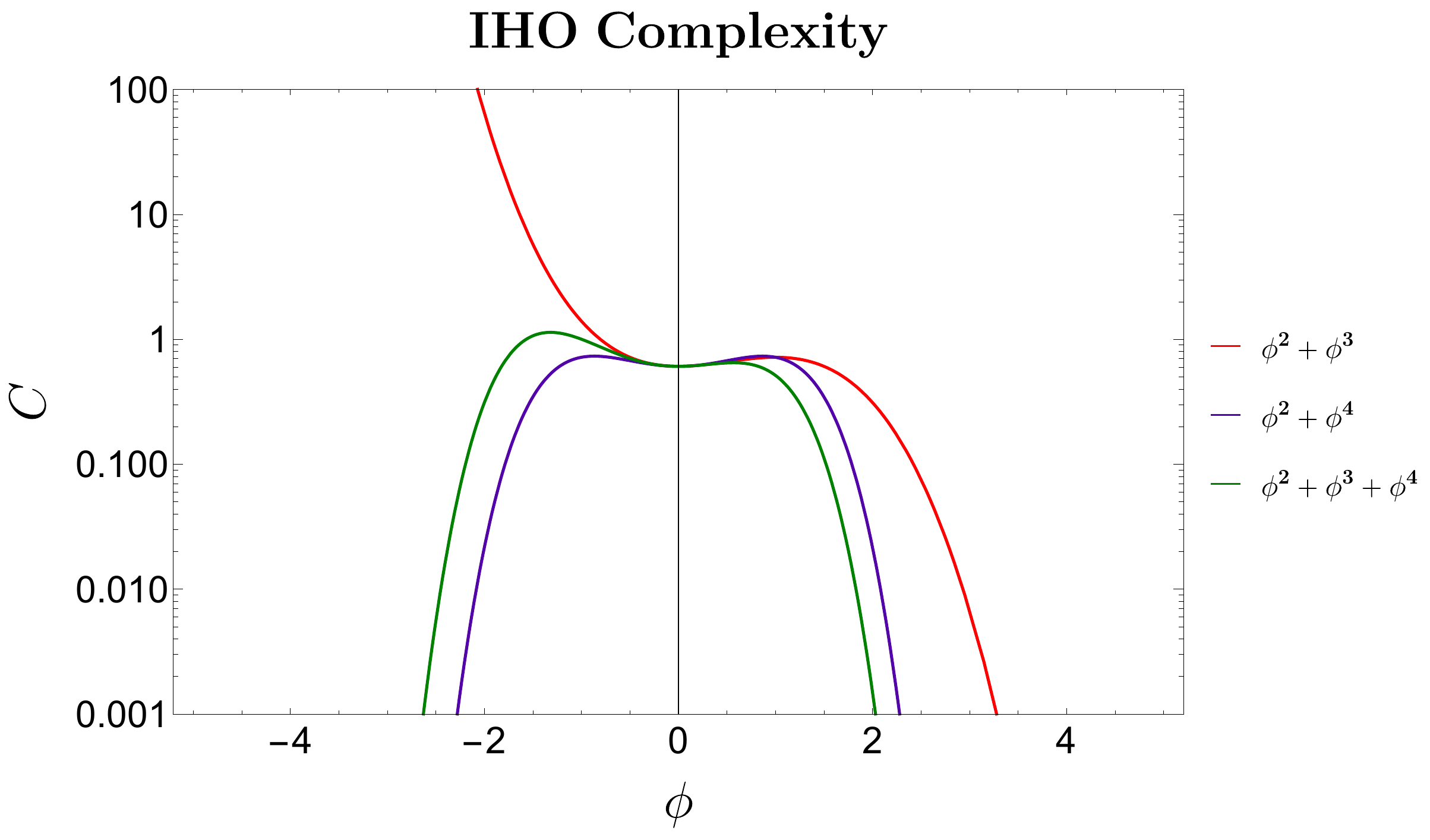}	
	\caption{Complexity against $\phi$ for different perturbations of field with $\lambda=-1$}\label{fig:IHOneg}
\end{figure}
\begin{figure}[!htb]
	\centering
	\subfigure[][Variation of inverse temperature]{\includegraphics[height=9cm,width=17cm]{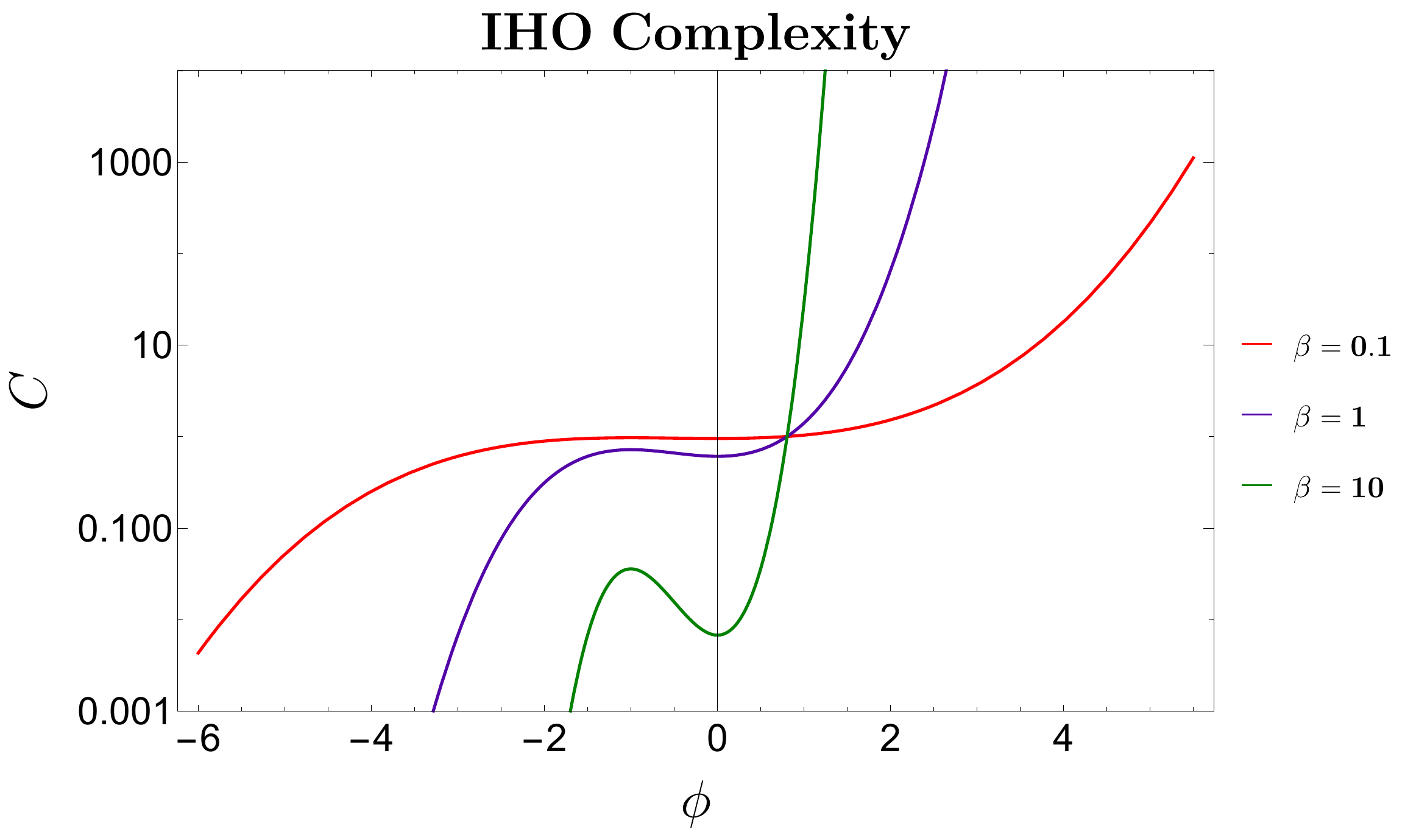}\label{fig:IHO3b}}\hfill
	\subfigure[][Variation of mass]{\includegraphics[height=9cm,width=17cm]{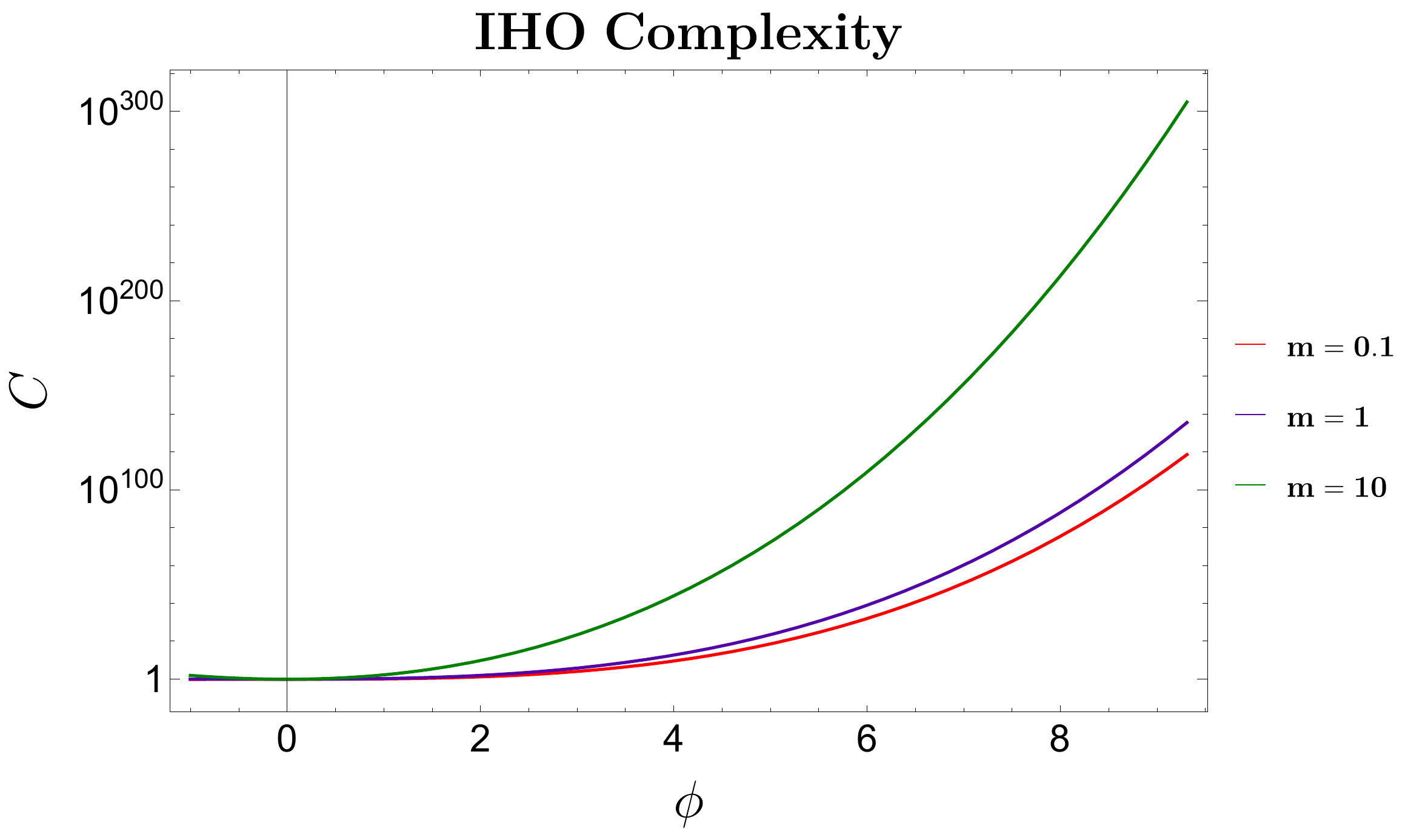}\label{fig:IHO3m}}
	\caption{behvior of the complexity of $\phi^2 + \phi^3$ perturbation with varying $\beta$ and $m$}
	\label{fig:IHO3bm}
\end{figure}
\begin{figure}[!htb]
	\centering
	\subfigure[][Variation of inverse temperature]{\includegraphics[height=9cm,width=17cm]{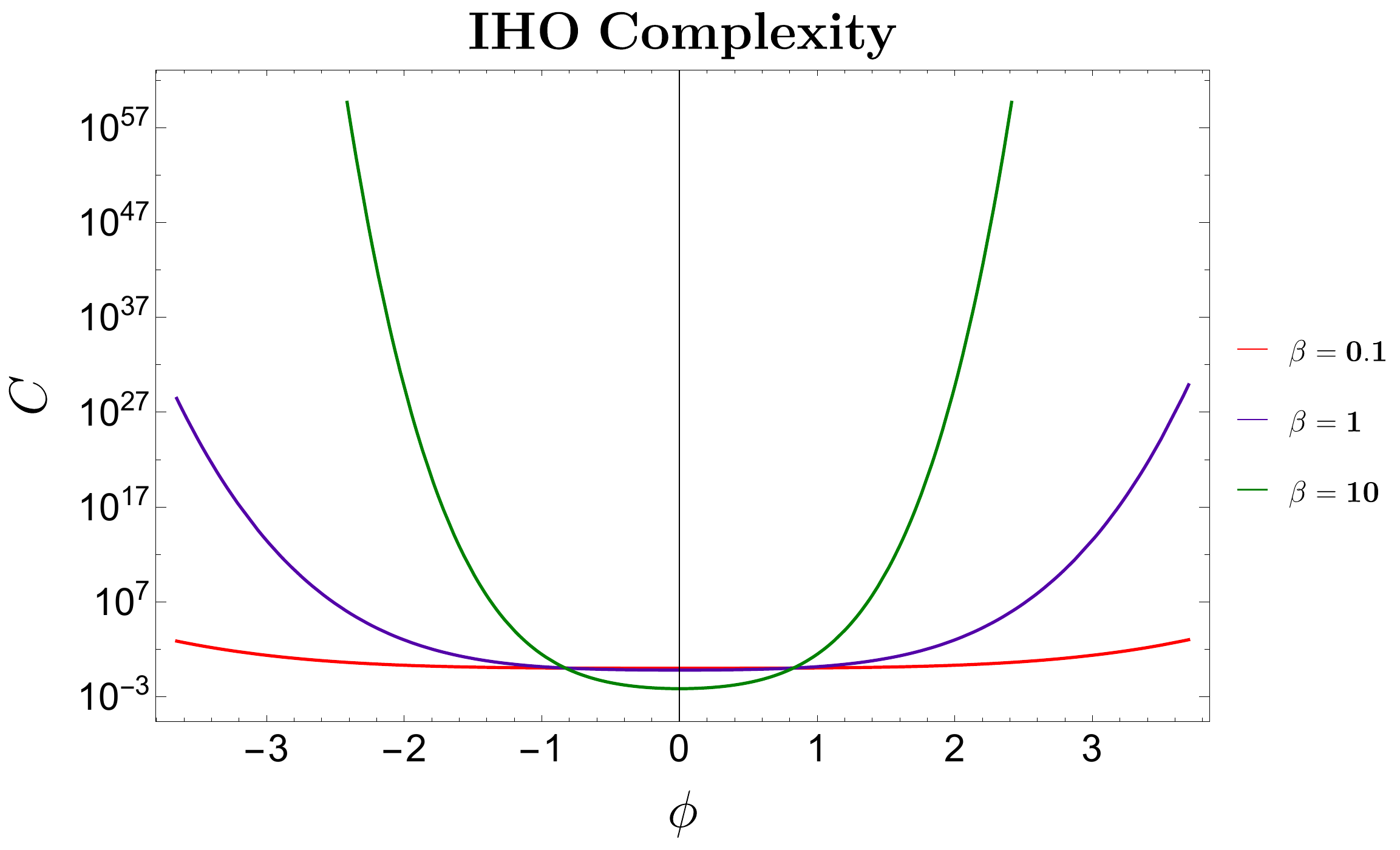}\label{fig:IHO4b}}\hfill
	\subfigure[][Variation of mass]{\includegraphics[height=9cm,width=17cm]{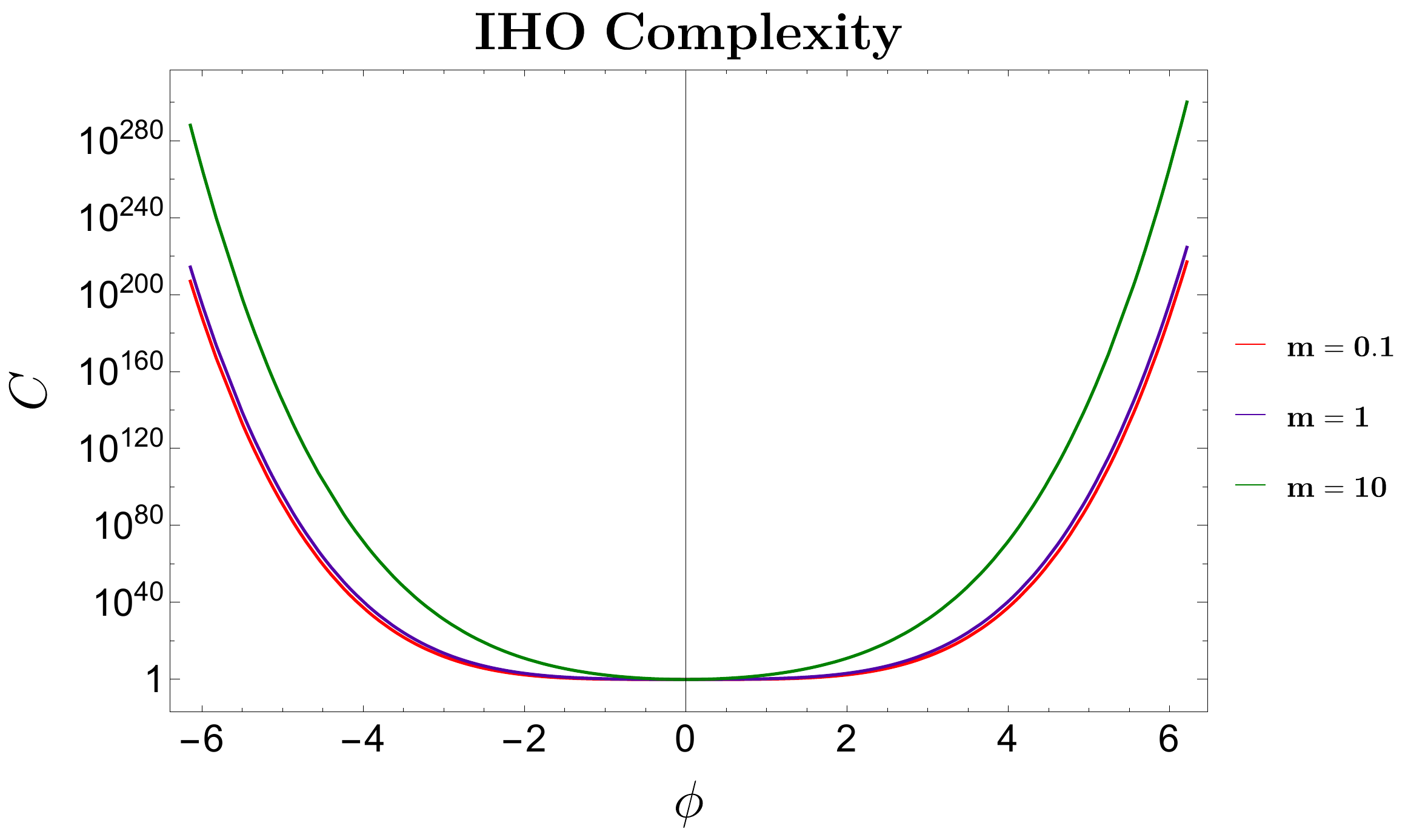}\label{fig:IHO4m}}
	\caption{behvior of the complexity of $\phi^2 + \phi^4$ perturbation with varying $\beta$ and $m$}
	\label{fig:IHO4bm}
\end{figure}

Before we proceed to vary the inverse temperature ($\beta$) and the mass ($m$), we can take a look at the general feature of the graphs of different perturbations in the field for both positive and negative $\lambda$ as we have done in \Cref{fig:IHO} and \Cref{fig:IHOneg}.

We see that the complexity values in all perturbations rise for positive coupling coefficient but at different rates. In contrast, in the case of a negative coupling coefficient, we see that they decrease rapidly after an initial period of rising (almost an inverse behaviour of the positive $\lambda$ case). We also observe symmetry for negative values of the field in the $\phi^2 + \phi^4$ graph. It is important to note that although we can calculate the slope values, one cannot associate a \textit{Lyapunov exponent} for the same because saturation does not exist. Maybe we can regard the slight rise and saturation before the dip in the case of negative Lambda values and associate a \textit{Lyapunov exponent} to it.

In \Cref{fig:IHO3bm}, we have plotted how the complexity values vary when we vary the inverse temperature and the mass. Very similar behaviour is observed in the case of $\phi^2 +\phi^4$ perturbation, shown in \Cref{fig:IHO4bm}.

Since we have already established the behaviour of the complexity in the case of negative $\lambda$, we take the liberty of not explicitly plotting the above variations in the inverse temperature and mass for that case. One has to imagine a decreasing (cannot associate a Lyapunov exponent) mirror inversion in the negatively valued domain of the field. This is true for the cases of both perturbations that we have discussed in the present context. 


\textcolor{Sepia}{\section{\sffamily Quantum Chaos from Morse function}\label{sec:10}} 

The emergence of chaos in quantum phenomena can be estimated using an out-of-time-order correlation function that is firmly associated with the operator commutator, split up in time. However, the universal relation $\displaystyle \mathcal{C} = -\ln(OTOC)$ relating complexity with OTOC \cite{Choudhury:2020hil, Bhargava:2020fhl, Hashimoto:2017oit, Choudhury:2020yaa, Choudhury:2018fpj, Choudhury:2018rjl,Choudhury_2021c,BenTov:2021jsf} has been studied greatly in recent times to diagnose chaos in various physical models. This section will prove this universal relation for the supersymmetric case, relating the complexity with OTOC using the frame of Morse theory. We will comment on the upper bound of chaos, namely the Lyapunov exponent. In the above sections, by identifying the Morse function on a manifold with the cost function, we have calculated the complexity for supersymmetric field theories in various regimes in terms of the Hessian $H(f)$ and also have made use of the fact that the eigenvalues of supersymmetric Hamiltonian $H_{t}$ are concentrated near the critical points of the Morse function defined on the manifold. In this section, we will make use of the same facts and derive the universality relation. 
\\
\par{\textcolor{Sepia}{\bf\sffamily Theorem} - \textit{Let $\displaystyle \phi(t)$ be an integral curve which represents the state of the particle at various time, then if $\displaystyle x_{1},...x_{m}$ be a local coordinate chart around a critical point $\displaystyle p \in M$ such that $\displaystyle\frac{\partial}{\partial x^{1}},....,\frac{\partial}{\partial x^{m}}$ is an orthonormal basis for $\displaystyle T_{p}M$ with respect to the metric $\displaystyle g$, then for any $\displaystyle t\in R$ the out-of-time-correlator at $\displaystyle p$ is equal to the exponential of the minus the matrix of the Hessian at $\displaystyle p$, i. e.}}
\begin{equation}
\large \frac{\partial}{\partial x}\phi_{t}|_{p} = \exp(-H_{p}(f)t).
\end{equation}
By identifying $\phi_{t}$ as a smooth function on the surface of the manifold we have
\begin{equation}
\large \frac{d}{dt}\phi(t, x) = -(\nabla f)\left(\frac{\partial}{\partial x}\phi(t, x)\right).
\end{equation}

By changing the order of differentiation for any $x \in M $ we get,
\begin{equation}
\large \frac{d}{dt}\left(\frac{\partial}{\partial x}\phi(t, x)\right) = -\left(\frac{\partial}{\partial x}\nabla f\right)\left(\frac{\partial}{\partial x}\phi(t, x)\right).
\end{equation}
Therefore by defining: 
\begin{equation}
\large \Phi(t, x) = \frac{\partial}{\partial x}\phi(t, x)
\end{equation} 
is a solution of the linear system of ODE's:
\begin{equation}
\large \frac{d}{dt}\Phi(t, x) = -\left(\frac{\partial}{\partial x}\nabla f\right)\Phi(t, x) .
\end{equation}
Because $\displaystyle \exp\left(-\left(\frac{\partial}{\partial x}\nabla f\right)t\right)$ is also a solution to the linear ODE's, we finally get:
\begin{equation}
\large \Phi(t, x) = \frac{\partial}{\partial x}\phi(t, x)=\exp\left(-\left(\frac{\partial}{\partial x}\nabla f\right)t\right).
\end{equation}
Since the solution is unique, thereby, at the location of critical point $p$ we have:
\begin{equation}
\large \frac{\partial}{\partial x}\phi_{t}|_{p} =\exp(-H_{p}(f)t)
\end{equation}
 
Hence by identifying the complexity of SUSY field theories with the Hessian as described in this paper, the above equation under good approximations could be written as:
\begin{equation}
\large \mathcal{C} = -\ln(OTOC).
\end{equation}

From the above equation, we could see that the Hessian of the Morse function, under good approximations, is perfectly consistent with \textit{universal relation} and is of great interest to capture the effect of chaos in supersymmetric field theories. 

\par{we will now comment on the behaviour of Lyapunov exponent in SUSY field theories, especially in the framework of an inverted harmonic oscillator under which it is expected to have chaotic features, depending upon the increase in the number of critical points as described in section \ref{sec:6}. Thus we could write the expression for complexity in the IHO regime as:}
\begin{equation}
\large \mathcal{C}_{i}(t) \approx c~\exp{(\lambda_{i} t)} \ \ \forall \ \ i = 1, 2,...,n.
\end{equation}

It is to be noted that the above equation is valid only for the IHO. The index $i$ indicates the higher order quantum corrections in the Lagrangian for which the complexity has been measured, then mathematically Lyapunov exponent could be written as:
\begin{equation}
\large \lambda_{i} = \left(\frac{d \ln\mathcal{C}_{i}(t)}{dt}\right)\ \ \forall \ \ i = 1, 2,...n.
\end{equation}

Now using the universal relation relating complexity with the Out-of-Time Ordered Correlation (OTOC) function, one could write:
\begin{equation}
\large OTOC = \exp(-c\exp{(\lambda t)})
\end{equation}
where $\lambda$ is the Lyapunov exponent \cite{Maldacena:2015waa, Bunakov:2017ghk, Han:2018bqy} which captures the effect of chaos in the quantum regime and relates different measures of complexity with OTOC through the universal relation. The above universal relation between complexities in different order of perturbations in the superpotential can be translated to the Lyapunov exponent through the MSS bound as:
\begin{equation}
\large \lambda_{i} \le \lambda \le \frac{2\pi}{\beta} \ \ \forall \ \ i = 1,2,...,n.
\end{equation}
where $\beta$ is the inverse temperature. We have shown that the Morse function could be used to classify the topology of surfaces and capture the effect of chaos in the quantum regime of supersymmetric field theories.

\textcolor{Sepia}{\section{\sffamily Conclusions}\label{sec:11}}
Out-of-order-correlation-function (OTOC) in the framework of supersymmetry has been studied before and, as a result, didn't show any chaotic behaviour in the regime of SHO. Our primary focus in this paper was to bring out the relationship between circuit complexity and Morse function and comment on the complexity of supersymmetric quantum field theory, in the regime of the simple and inverted harmonic oscillator (IHO), by formulating the potential of IHO as the generators of $SL(2, Z)$ group. By pointing out the relationship between the cost and Morse function on a manifold, we have explicitly shown how the critical points on the surface encapsulate the action of the supersymmetric charge on the given reference state. We have explicitly made use of the fact that the eigenvalues of the supersymmetric Hamiltonian are concentrated near the critical points of the Morse function in the manifold, and then using the Witten index, which comments on the symmetry breaking of the theory, we commented on the complexity of the supersymmetric field theories for the IHO, which increases by a factor of exponential. For computations of complexity in the regime of the simple harmonic oscillator, we found out that circuit complexity didn't show any dependence on initial conditions or exponential behaviour. Next, we have the well known universal relation relating complexity and out-of-time ordered correlation function $\mathcal{C} = -ln(OTOC)$ using the general description of Morse theory. 

\begin{itemize}
	\item \underline{\textcolor{red}{\bf Remark~I:}}\\The circuit complexity for supersymmetric field theories has very deep connections with the Morse function defined on the surface of the manifold. We have obtained the expression for complexity for SUSY field theories in terms of the Hessian of the Morse function. In doing so, we have made use of the fact the eigenvalues of the supersymmetric Hamiltonian are concentrated near the critical points of the Morse function. 
	
	\item \underline{\textcolor{red}{\bf Remark~II:}}\\ The behaviour of complexity for supersymmetric field theories for the inverted harmonic oscillator is of prime importance, we have found that the growth of complexity in the regime of IHO is directly related to the growth of the number of critical points on the manifold, which in turn grows exponentially with respect to the superpotential, we observed similar behaviour for higher-order quantum corrections namely $\phi^{3}$ and $\phi^{4}$ theories. 
	
	\item \underline{\textcolor{red}{\bf Remark~III:}}\\ We have also computed complexity for the simple harmonic oscillator, and found out that circuit complexity didn't show any dependence on initial conditions or exponential behaviour. It is also worth mentioning that complexity for supersymmetric scalar field theories only depends on the absolute value of the non-dynamical auxiliary field. The $F$ - term is identified with the gradient of the Morse function determines whether the supersymmetry is spontaneously broken or not depending upon whether the gradient has passed through the critical points. On passing through the critical points, we get $F$ = 0 which means no zero energy supersymmetric ground states exist. \\

	\item \underline{\textcolor{red}{\bf Remark~IV:}}\\ We have proved the well known universal relation $\mathcal{C} = -ln(OTOC)$ which relates complexity with the out-of-time ordered correlation function for supersymmetric field theories using Morse theory. The out-of-time ordered correlation function is an excellent gadget to capture the effect of chaos in the quantum regime. In this paper, we have obtained an upper bound on the Lyapunov exponent and also commented on its various features for supersymmetric field theories purely for SHO and IHO in table \ref{table:a} and \ref{table:b} using aspects of Morse Theory. The main point of Witten's paper on supersymmetry and Morse theory was to provide supersymmetry with a mathematical structure. Like the Witten index, which tells if the supersymmetry is broken or not, various results wouldn't have been possible by the normal description of particle physics. \\ 
	
	\item \underline{\textcolor{red}{\bf Remark~V:}}\\ We have found that complexity for supersymmetric field theories differ significantly from ordinary quantum field theories in the sense that for non-SUSY QFT complexity, slowly starts to increase at the critical point, however for SUSY field theories the graph rises fast initially and then gradually tends to saturate. The rate of saturation depends on the order of quantum corrections. The $\phi^{3}$ theory saturates very rapidly and have the smallest Lyapunov exponent among the other studied perturbations, while the complexity for theory involving $\phi^{4}$ term saturates slowly. For IHO, we observed that complexity increases exponentially and quickly rose to very high values unlike ordinary QFT where it has a linear growth.\\
	
	\item \underline{\textcolor{red}{\bf Remark~VI:}}\\ We have explicitly studied the dependence of mass on the behaviour of circuit complexity and have observed that for massive fields there is a decrease in the rate of change of complexity for $\phi^{4}$ theory, and it interesting to note that the graph becomes nearly indistinguishable from that of free field theory. However, for smaller masses, the complexity for $\phi^{3}$ graph approaches to $\phi^{2}$ slowly with a slight decrease in the rate of complexity. Hence, we expect an increase in the value of the Lyapunov exponent. In the case of IHO, we observe that as mass increases, the rate of change of complexity w.r.t $\phi$ also increases.\\
	
	\item \underline{\textcolor{red}{\bf Remark~VII:}}\\ We have also commented on the behaviour of complexity w.r.t the coupling constant $\lambda$ and have observed that for negatives values of $\lambda$ in the regime of SHO the saturation is slower for $\phi^{2}$ and $\phi^{3}$ perturbations and in the case of $\phi^{4}$ theory their is a sudden dip at the initial stage to zero thereby right shifting the point of initial rise of complexity. In the case of IHO, the increase in the value of $\lambda$ results in an increase in the rate of change complexity; however, for negative values of $\lambda$ we found that complexity decreases exponentially SUSY field theories.

\end{itemize}

The future prospects of the present work are appended below:
\begin{itemize}
	\item \underline{\textcolor{red}{\bf Prospect~I:}}\\ In this paper we have restricted ourselves to supersymmetric scalar field theories, however similar computations could be done for supersymmetric gauge and non-abelian gauge theories by taking in consideration the dynamical $D$ - term   ~\cite{Itoyama:2013sn, Piguet:1997vg}, which would give further understanding about complexity and effect of chaos in supersymmetric field theories.\\
	
\item \underline{\textcolor{red}{\bf Prospect~II:}} circuit complexity for interacting quantum field theories and its relation with renormalization group has been studied by Arpan Bhattacharyya and collaborator and thus, it will be interesting to see what new mathematical structure does renormalization group flow brings out and how it is related to complexity\cite{Bonini:1998ec, Kazakov:2000us, Maruyoshi:2016tqk}.\\
	
	\item \underline{\textcolor{red}{\bf Prospect~III:}} In this paper, we have computed the complexity for SUSY scalar field theories by making use of the properties of the Hessian matrix. However, we hope that this is not all\cite{Murugan:2017eto}. The use of other remarkable properties of the Morse function could help in gaining a much broader perspective in supersymmetric field theories and its matter content and interactions and the effect it has on the expansion of universe. \cite{Bilic:2010xd}. \\
	
	\item \underline{\textcolor{red}{\bf Prospect~IV:}} The study of supersymmetry and its complexity in terms of Morse theory has given it a geometrical structure, however for theories of supergravity it is still not quit clear what the right mathematical structure is\cite{Coimbra:2011nw, Lazaroiu:2020vne}, we suppose that the work on this direction would bring interesting connections between quantum chaos and various other mathematical theories.\\
	
	\end{itemize}

	\subsection*{Acknowledgements}
	The research fellowship of SC is supported by the J. C. Bose National Fellowship of Sudhakar Panda. Also SC take this opportunity to thank sincerely to Sudhakar Panda for his constant support and providing huge inspiration. SC also would line to thank School of Physical Sciences, National Institute for Science Education and Research (NISER), Bhubaneswar for providing the work friendly environment. SC also thank all the members of our newly formed virtual international non-profit consortium ``Quantum Structures of the Space-Time \& Matter" (QASTM) for elaborative discussions. Satyaki Choudhury, Sachin Panner Selvam and K. Shirish would like to thank NISER Bhubaneswar, BITS Hyderabad, VNIT Nagpur respectively for providing fellowships. Last but not the least, we would like to acknowledge our debt to 
the people belonging to the various part of the world for their generous and steady support for research in natural sciences. 
	
\clearpage
	\appendix

\newpage
\bibliographystyle{utphys}
\bibliography{bibliography.bib}

\end{document}